\documentclass[12pt]{article}
\usepackage[utf8]{inputenc}
\usepackage[british]{babel}
\usepackage{cmap}
\usepackage{lmodern}
\usepackage[T1]{fontenc}

\usepackage{amssymb, amsmath, amsthm}
\usepackage[a4paper,top=25mm,bottom=25mm,left=25mm,right=25mm]{geometry}
\usepackage{ragged2e}

\usepackage{authblk} 
\usepackage{pifont}
\usepackage{graphicx}
\usepackage[dvipsnames,svgnames,table]{xcolor}
\usepackage[figuresright]{rotating}
\usepackage{xtab} 
\usepackage{longtable} 
\usepackage{multirow}
\usepackage{footnote}
\usepackage[stable]{footmisc}
\usepackage{chngpage} 
\usepackage{pdflscape} 
\usepackage[nottoc,notlot,notlof]{tocbibind} 

\usepackage{pgfplots}
\pgfplotsset{every tick label/.append style={font=\footnotesize}}
\pgfplotsset{compat=1.18}
\usepackage{setspace}

\usepackage{array}
\newcolumntype{K}[1]{>{\centering\arraybackslash$}p{#1}<{$}}

\makesavenoteenv{tabular}
\usepackage{tabularx}
\usepackage{booktabs}
\usepackage{threeparttable}
\usepackage[referable]{threeparttablex} 
\newcolumntype{R}{>{\raggedleft\arraybackslash}X}
\newcolumntype{L}{>{\raggedright\arraybackslash}X}
\newcolumntype{C}{>{\centering\arraybackslash}X}
\newcolumntype{A}{>{\columncolor{gray!25}}C}
\newcolumntype{a}{>{\columncolor{gray!25}}c}

\newlength{\tablen}

\usepackage{dcolumn} 
\newcolumntype{.}{D{.}{.}{-1}}

\usepackage{tikz}
\usetikzlibrary{arrows, calc, matrix, patterns, positioning, trees}
\usepackage[semicolon]{natbib} 
\usepackage[hyphens]{xurl}
\usepackage{microtype}
\usepackage[justification=centering]{caption} 

\usepackage[labelformat=simple]{subcaption}

\DeclareCaptionLabelFormat{parenthesis}{(#2)}
\makeatletter
\renewcommand\p@subfigure{\arabic{figure}.}
\makeatother

\DeclareCaptionLabelFormat{parenthesis}{(#2)}
\makeatletter
\renewcommand\p@subtable{\arabic{table}.}
\makeatother

\usepackage{enumitem}

\setlist[itemize]{leftmargin=2.5\parindent}
\setlist[enumerate]{leftmargin=2.5\parindent}

\usepackage{hyperref} 
\hypersetup{
  colorlinks   = true,    		
  urlcolor     = blue,    		
  linkcolor    = blue,    		
  citecolor    = ForestGreen	
}

\def\addlegendimage{\csname pgfplots@addlegendimage\endcsname}

\theoremstyle{plain}

\newtheorem{lemma}{Lemma}
\newtheorem{proposition}{Proposition}
\newtheorem{theorem}{Theorem}

\theoremstyle{definition}

\newtheorem{definition}{Definition}[section]
\newtheorem{example}{Example}

\theoremstyle{remark}

\makeatletter
\let\@fnsymbol\@alph
\makeatother

\def\keywords{\vspace{.5em} 
{\noindent \textit{Keywords}: }}

\def\AMS{\vspace{.5em} 
{\noindent \textbf{\emph{MSC} class}: }}

\def\JEL{\vspace{.5em} 
{\noindent \textbf{\emph{JEL} classification number}: }}

\title{Reducing the non-uniformity of the \\ group draw in sports tournaments}
\author{\href{https://sites.google.com/view/laszlocsato}{L\'aszl\'o Csat\'o}\thanks{~E-mail: \emph{laszlo.csato@sztaki.hun-ren.hu}} 
}
\affil{Institute for Computer Science and Control (SZTAKI) \\
Hungarian Research Network (HUN-REN) \\
Laboratory on Engineering and Management Intelligence \\
Research Group of Operations Research and Decision Systems}
\affil{Corvinus University of Budapest (BCE) \\
Institute of Operations and Decision Sciences \\
Department of Operations Research and Actuarial Sciences}
\affil{Budapest, Hungary}
\date{\today}

\def\Dedication{
{\noindent
``\emph{Plus on vieillit, et plus on se persuade que Sa sacr\'ee Majest\'e le Hasard fait les trois quarts de la besogne de ce mis\'erable univers.}''\footnote{~``\emph{The older we get, the more we convince ourselves that His sacred Majesty Chance does three quarters of the work of this miserable universe.}'' (Source: Correspondance de Voltaire, 1759, Lettre 3803. \url{https://fr.wikisource.org/wiki/Correspondance_de_Voltaire/1759/Lettre_3803})}
}
\vspace{0.25cm}

\flushright
\noindent (Frederick the Great, the King of Prussia, to Voltaire)

\vspace{1cm} 
\justify }

\begin{document}

\newgeometry{top=10mm,bottom=15mm,left=25mm,right=25mm}

\maketitle
\Dedication

\begin{abstract}
\noindent
The group draw of a sports tournament requires assigning teams to groups of (almost) the same size. The most important criteria for a draw procedure are balance, randomness, and transparency, which could not be satisfied simultaneously if draw constraints exist. Organisers usually use the so-called Skip mechanism, a method based on a random sequential draw of the teams from pots, in order to ensure balance and transparency. However, the Skip mechanism is non-uniformly distributed: the valid assignments are not necessarily equally likely. We quantify this distortion if a group can contain at most two teams from a given set S, which poses a serious challenge for the Skip mechanism. Our study provides exact results for an arbitrary number of teams when there are three pots and two pots contain only one team from the set S, as well as complete enumeration for small problems with three pots and at most five teams per pot. We also analyse three real-world case studies from basketball and football.
It turns out that the optimal design considers the pots in decreasing order according to the number of teams in the set S.
These results can be used to identify the least distorted transparent draw procedure, and decide whether the extent of non-uniformity calls for further actions.


\AMS{62-08, 90-10, 90B90, 91B14}

\JEL{C44, C63, Z20}
\end{abstract}

\clearpage
\restoregeometry

\section{Introduction} \label{Sec1}


In the group draw of a sports tournament, the teams are assigned to some groups where a round-robin contest is played to determine the set of teams qualifying for the next stage. The allocation is made by a draw procedure. The most important criteria to evaluate such a mechanism are \emph{balance}, \emph{randomness}, and \emph{transparency} \citep{Guyon2015a}.

According to balance, the expected strength of the groups needs to be roughly equal \citep{LalienaLopez2019, LalienaLopez2025}. In the absence of balance, a strong team may have to play against strong opponents in the group stage, making qualification less likely than for a weaker team, which plays against weak opponents in the group stage.
Randomness prevents accusations that the organiser manipulates the draw for the sake of some favoured teams \citep{RobertsRosenthal2024}.
Finally, the randomisation should be \emph{perceived} as fair by all stakeholders, which can be called transparency \citep{BoczonWilson2023}. In other words, it is not sufficient if randomness is guaranteed by a black box algorithm since the outcome of the draw will inevitably benefit some teams at the expense of other teams when the fans of the latter may suspect an unfair intervention by the organiser.

In practice, the balancedness of the groups is usually guaranteed by dividing the teams into pots according to their strength based on historical performances. The number of pots equals the number of teams to be assigned to a group; the first pot contains the strongest teams, the second pot contains the next teams in the ranking, and so on.
The three criteria can be satisfied by drawing the teams sequentially from the pots, and placing them in the first available group in alphabetic order such that each group contains one team from each pot. 

However, draw constraints might make some allocations invalid. For instance, the FIFA World Cup---the most prestigious football competition, and one of the most widely viewed sporting event in the world---maximises the number of intercontinental games in the group stage by restricting the number of teams from the same geographic zone in a group. The Union of European Football Associations (UEFA) apply different draw conditions in its competitions ``\emph{to issue a schedule that is fair for the participating teams, fulfils the expectations of commercial partners and ensures with a high degree of probability that the fixture can take place as scheduled}'' \citep{UEFA2020c}.
Draw restrictions can also be used to decrease the probability of matches with misaligned incentives \citep{Csato2022a}, or to avoid certain games due to security reasons \citep{Kobierecki2022}.

In the presence of draw constraints, it becomes challenging to satisfy the three criteria.
Balance cannot be compromised to ensure competitive balance, which is essential for an exciting tournament \citep{LapreAmato2025, LaprePalazzolo2023}. Therefore, the trade-off between randomness and transparency becomes a key issue for the organiser.

Draw ceremonies are attended by many celebrities and thousands of guests, streamed live over the internet, and broadcast by media companies \citep{BoczonWilson2023}. All stakeholders and, especially, fans should be persuaded that their teams are treated fairly and the composition of the groups is not decided by bargaining behind closed doors or by opaque---and potentially biased---computer programs. These arguments call for transparency: the random components of the draw should be credible and easy to comprehend, while its deterministic parts remain verifiable during the draw with basic mathematical knowledge.

\subsection{The research question and its managerial relevance} \label{Sec11}

The most popular balanced and transparent draw procedure in sports is the so-called \emph{Skip mechanism}. This method is currently used by at least three governing bodies, the FIBA (F\'ed\'eration internationale de basketball, International Basketball Federation), the FIFA (F\'ed\'eration internationale de football association, International Association Football Federation), and the UEFA in various tournaments, including the FIBA Basketball World Cup, the FIFA World Cup, and the UEFA Nations League \citep{Csato2025f}.

However, the Skip mechanism does not satisfy randomness: it is non-uniform over the set of feasible assignments, the valid allocations do not have the same chance to occur \citep{Csato2025c, RobertsRosenthal2024}. For instance, in the 2018 FIFA World Cup draw, the Skip mechanism distorted the probability of qualification by more than 0.5 percentage points for seven national teams \citep{Csato2025c}. Analogously, the 2022 FIFA World Cup draw increased the probability that the United States and Qatar play in the same group by 38\%, from about 9.06\/ to 12.5\% \citep[p.~665]{RobertsRosenthal2024}.

Since no draw procedure is known to satisfy both randomness and transparency \citep{RobertsRosenthal2024}, it is crucial to uncover the extent to which the Skip mechanism is non-uniform.
Furthermore, the Skip mechanism is essentially a family of draw procedures because the assignment probabilities depend on the \emph{draw order}, on the sequence in which the pots are emptied \citep{Csato2025c}. Using the Skip mechanism with an optimal draw order is certainly easier to implement than adopting a completely new method. Even if no reasonable definition of transparency exists in the literature, changing the draw order is obviously less threatening for transparency than a more fundamental reform in the draw procedure.

Therefore, we will address two issues:
(a) What is the effect of the draw order on the bias of the Skip mechanism?
(b) What is the optimal order of the pots?

\subsection{An overview of the results} \label{Sec12}

Section~\ref{Sec3} investigates two different types of draw constraints that are widely used in practice. One set of constraints from the first type does not make the Skip mechanism biased (Lemma~\ref{Lemma1}), but this is not true for the second type of constraints (Lemma~\ref{Lemma2}).

Therefore, Section~\ref{Sec4} focuses on the second type of constraints: from an exogenously given set $S$, no more than two teams are allowed to be assigned to the same group.
The distortions are computed analytically for any number of teams in Section~\ref{Sec41}. This remains possible only if there are three pots and the set $S$ contains only one team from two pots. Further extension of exact results seems to be excluded by the rapidly increasing complexity of the problem. Hence, Section~\ref{Sec42} numerically computes the aggregated distortions for all possible configurations of $S$ with three pots and at most five groups.

The conclusions point in the same direction: the best design is a decreasing order of the pots according to the number of teams from the set $S$. The advantage of the optimal draw order turns out to be substantial, often reaching 75--80\% in relative terms (Table~\ref{Table2}).

These findings are reinforced by three real-world case studies in Section~\ref{Sec5}. All these examples impose a draw constraint similar to the theoretical model (no group can contain more than two teams from a given set), besides several further restrictions.
Section~\ref{Sec51} returns to the aggregated distortions of the $4! = 24$ possible draw orders for the 2018 FIFA World Cup draw and \emph{explains} the biases reported by \citet{Csato2025c} in view of the analysis in Sections~\ref{Sec3} and \ref{Sec4}.
Section~\ref{Sec52} compares 11 selected draw orders---out of the $6! = 720$ options---for the European Qualifiers for the 2022 FIFA World Cup.
Finally, Section~\ref{Sec53} gives a comprehensive analysis of the 2019 FIBA Basketball World Cup draw. In contrast to the previous two practical examples, the assignment probabilities can be exactly computed here due to the manageable number of feasible solutions. It turns out that some draw orders can be equivalent to a uniform draw, while others remain (strongly) biased: the distortion of the assignment probability can exceed 15 percentage points for certain pairs of teams. 

\subsection{Main contributions} \label{Sec13}

Our paper extends the knowledge about group draw procedures in three directions.

First, we provide \emph{theoretical} results on the distortions of a field-proven randomisation procedure used in the group draw of sports tournaments, which has never been done before. Even though the UEFA Champions League Round of 16 draw \citep{BoczonWilson2023, KlossnerBecker2013}, the UEFA Champions League league phase draw \citep{GuyonBenSalemBuchholtzerTanre2025}, and the FIFA World Cup draw \citep{Csato2025c, Guyon2015a, RobertsRosenthal2024} have been extensively analysed, previous works neither derived analytical formulas for the distortions, nor provided a comprehensive study of small problems.

Second, the importance of draw order is highlighted. According to previous studies, the draw order has only a marginal effect on the alternative Drop mechanism in the UEFA Champions League Round of 16 draw \citep{BoczonWilson2023, KlossnerBecker2013} and in the UEFA Champions League league phase draw \citep{GuyonBenSalemBuchholtzerTanre2025}. Although \citet{Csato2025c} has already presented the non-negligible impact of draw order in the 2018 FIFA World Cup draw, this is only one simulated case study, and the results will be explained in Section~\ref{Sec51} of the current paper.

Third, the existing literature focuses on guaranteeing uniform distribution for the group draw by a novel---but less transparent---draw procedure \citep{Guyon2014a, KlossnerBecker2013, RobertsRosenthal2024} rather than amending the expected outcome by slightly changing an existing and already used mechanism without compromising its full transparency. The latter option may be more acceptable for tournament organisers and has a higher chance of being implemented in practice, even if it does not promise to eliminate the distortions.

To summarise, the findings presented here yield a clear policy recommendation for optimising several real-world group draws (e.g.\ FIBA Basketball World Cup draw, FIFA World Cup draw, UEFA Nations League draw) in order to reduce their non-uniformity gap.

\section{Related literature} \label{Sec2}

Fairness of sports rules is an extensively researched topic in operational research, see, for example, \citet{Csato2021a, KondratevIanovskiNesterov2024, KendallLenten2017, LentenKendall2022}. A recent survey overviews the fairness of draw rules in sports tournaments \citep{DevriesereCsatoGoossens2025}.

Inspired by the format of the FIFA World Cup, several studies have developed draw systems for competitions with geographical restrictions to create balanced groups at roughly the same competitive level \citep{CeaDuranGuajardoSureSiebertZamorano2020, Guyon2015a, LalienaLopez2019, LalienaLopez2025}. \citet{Csato2023d} demonstrates that the 2022 FIFA World Cup draw has not balanced the groups to the extent possible. The draw procedures of the 1990 \citep{Jones1990}, 2006 \citep{RathgeberRathgeber2007}, 2014 \citep{Guyon2015a}, 2018 \citep{Csato2025c} and 2022 FIFA World Cups \citep{RobertsRosenthal2024} have been verified to violate randomness. \citep{Csato2025c} and \citet{RobertsRosenthal2024} explain how the Skip mechanism can be simulated. \citet{RobertsRosenthal2024} propose some uniformly distributed, albeit less transparent draw procedures.

The bias of another randomisation procedure, the Drop mechanism---used in UEFA club competitions---has also been quantified \citep{BoczonWilson2023, KlossnerBecker2013, Kiesl2013}. \citet{Kiesl2013} and \citet{WallaceHaigh2013} reveal the relation of the Drop mechanism and Hall's marriage theorem. Since the 2024/25 season, the Drop mechanism is used in the novel league phase draw \citep{GuyonBenSalemBuchholtzerTanre2025}, but the non-uniformity of this draw has not been quantified yet.

\citet{BoczonWilson2023} offer a comprehensive analysis of the Drop mechanism in the UEFA Champions League Round of 16 draw and show that it resembles the fairest possible lottery in this setting.
\citet{Csato2025f} compares the performance of the Drop and Skip mechanisms on bipartite graphs up to 16 nodes.
However, both works consider only two groups, where more complicated draw constraints, such as the one that will be in our focus, could not appear. Furthermore, they do not give any general theoretical result for a high number of teams and do not attempt to understand the role of the draw order. 

\citet{GoossensYiVanBulck2020} explore trade-offs between various fairness criteria in scheduling round-robin tournaments, while \citet{LiVanBulckGoossens2025} analyse fairness trade-offs in multi-league scheduling. In contrast to scheduling, we do not know about any analogous study regarding draw procedures. Our results could be crucial in this respect as they uncover the minimal amount of randomness (uniformity) that should be given up in order to retain the fully transparent Skip mechanism for a group draw.

\citet{AtefYektaBergmanDay2023} consider a similar problem, assigning workers to sets of predefined size under some constraints. Here, each worker reports their preferences on fellow workers, and the most common design goals are \emph{efficiency} (maximisation of social welfare) and \emph{stability} (Could a subset of players benefit from breaking away?). In our model, stability is not an issue because the organiser can enforce any solution. While the preferences of the teams regarding group assignment remain unknown, it can be reasonably assumed that the preferences are highly (or even perfectly) correlated: each team would like to play against the weakest opponents. Therefore, requiring randomness is, in a sense, equivalent to maximising the welfare of the worst-off team.

\section{The Skip mechanism and a challenging constraint} \label{Sec3}

The problem presented in the Introduction can be formalised as follows.

\begin{definition} \label{Def1}
\emph{Draw problem}:
$mn + \ell$ teams are partitioned into $m+1$ pots, where the first $m$ pots contain $n$ teams each and the last pot contains the remaining $\ell < n$ teams (if $\ell = 0$, the $m$ pots have $n$ teams each). The teams should be assigned to groups such that each group contains at most one team from each pot, and further draw constraints may apply.
\end{definition}

In any draw problem, the Skip mechanism is able to find a valid assignment if it exists.

\begin{definition} \label{Def2}
\emph{Skip mechanism}:
The pots are considered in a predetermined order, that is, the draw continues with the next pot only after one pot is emptied.
The teams are selected one at a time, uniformly at random, and assigned to the next available group in alphabetical order as indicated by the computer that checks whether a draw condition occurs or is anticipated to occur.
\end{definition}

Clearly, the set of groups available to a team depends not only on its own attributes but also on the attributes of the teams already drawn and the teams still to be drawn.

In the following analysis, teams are denoted by numbers and pots by capital letters.
Since we focus on the groups, the phrases two teams ``(cannot) play against each other'' and ``are (not allowed to be) assigned to the same group'' are synonyms.

\begin{example} \label{Examp1}
There are $m = 3$ pots with $n = 3$ teams each. Pot A consists of teams 1--3, Pot B consists of teams 4--6, and Pot C consists of teams 7--9. Two draw constraints exist:
(1) team 3 cannot play against team 9;
(2) at most two of teams 1, 4, 7 can be in the same group.
\end{example}

Example~\ref{Examp1} is used to illustrate the Skip mechanism.

\begin{example} \label{Examp2}
Consider Example~\ref{Examp1} with the Skip mechanism such that the teams are selected in their numerical order. After Pots A and B are emptied, the first group contains teams 1 and 4, the second group contains teams 2 and 5, while the third group contains teams 3 and 6. The draw continues with Pot C. Team 7 is drawn and assigned to the second group because of draw constraint (2). Team 8 is drawn and assigned to the third group because, even though team 8 can play against both teams 1 and 4 that are in the first group, team 9 cannot be placed in the third group due to draw constraint (1).
\end{example}

A video of the 2026 FIFA World Cup draw, which used the Skip mechanism, is available at \url{https://www.youtube.com/watch?v=9HX_tQBA-Iw}.

The two types of draw constraints in Example~\ref{Examp1} appear in real-world tournament draws as will be seen in Section~\ref{Sec5}.
Constraint (1) can be called a prohibited pair and generalised as follows.

\begin{definition} \label{Def3}
\emph{Set-based prohibited pair list}:
Some teams are of type $S$ and at most one team from type $S$ can be assigned to the same group.
\end{definition}

The number of prohibited pairs generated depends on the cardinality of set $S$ and the distribution of its teams across the pots, as teams from the same pot certainly do not play against each other. For example, if set $S$ contains two teams from pot A, two teams from pot B, and one team from pot C, then there are eight prohibited pairs.

One set-based prohibited pair list can be effectively treated by the Skip mechanism according to the following result.

\begin{lemma} \label{Lemma1}
Consider a draw problem with one set-based prohibited pair list. The Skip mechanism is equivalent to a uniform draw over all valid assignments, independently of the draw order of the pots.
\end{lemma}

\begin{proof}
Consider the teams from two arbitrary pots $u$ and $v$ such that set $U$ ($V$) contains the teams of pot $u$ ($v$) and $\left| U \right| = \left| V \right| = n$. Since teams in $U \cap S$ cannot play against teams in $V \cap S$, a particular team in $U \cap S$ has a probability of $1/ \left( n - \left| V \cap S \right| \right)$ to play against a particular team in $V \setminus \left( V \cap S \right)$, and a particular team in $V \cap S$ has a probability of $1/ \left( n - \left| U \cap S \right| \right)$ to play against a particular team in $U \setminus \left( U \cap S \right)$ under a uniform draw. However, these are also the probabilities under the Skip mechanism since it does not differentiate between the teams of $U \setminus \left( U \cap S \right)$ and $V \setminus \left( V \cap S \right)$ because the teams from any pot are selected uniformly at random.

An analogous argument finishes the proof if, without loss of generality, $\left| V \right| = \ell < n$.
\end{proof}

The Skip mechanism may be non-uniform if at least two set-based prohibited pair lists exist \citep[Proposition~1]{Csato2025f}. However, even in this case, it is far from clear whether the draw order could have a meaningful impact on the distortions as prohibited pairs are symmetric: if team $s$ cannot be assigned to the group of team $t$, then this relation holds in the other direction, too.

On the other hand, constraint (2), which can be called a prohibited triplet, poses a challenge to the Skip mechanism because it is ``short-sighted''.

\begin{lemma} \label{Lemma2}
Consider a draw problem with one prohibited triplet. The Skip mechanism may be non-uniform over all valid assignments.
\end{lemma}

\begin{proof}
It is sufficient to provide a counterexample. Consider Example~\ref{Examp1} with only the draw constraint (2). The labelling of the groups is arbitrary; hence, there exist $3! \cdot 3! = 36$ different assignments, among which $2! \cdot 2! = 4$ are invalid as constraint (2) is violated. Hence, the number of feasible allocations is 32.

Teams 1 and 4 play in the same group in $2! \cdot 2! \cdot 2 = 8$ assignments, which implies a chance of 1/4 under a uniform draw.

However, this probability is 1/3 if pot C is drawn last since the Skip mechanism is not able to take into account that team 7 cannot be assigned to a group if teams 1 and 4 are already there.
On the other hand, if pot C is not the last pot, then teams 1 and 4 will be in the same group with a probability of $1/3 \cdot 0 + 2/3 \cdot 1/3 = 2/9$.
\end{proof}

The notion of prohibited triplet can be generalised similarly to prohibited pair.

\begin{definition} \label{Def4}
\emph{Set-based prohibited triplet list}:
Some teams are of type $S$ and at most two teams from type $S$ can be assigned to the same group.
\end{definition}

Lemmata~\ref{Lemma1} and \ref{Lemma2} uncover that a set-based prohibited triplet list is more challenging for the Skip mechanism than a set-based prohibited pair list. Furthermore, the draw order of the pots becomes relevant if the draw problem contains set-based prohibited triplet lists. Therefore, we will focus on this type of constraint in the following. It has a strong practical relevance since both the FIFA World Cup and the European Qualifiers for the FIFA World Cup contain such a restriction, see Section~\ref{Sec5}.

\section{Theoretical findings for three pots} \label{Sec4}

Inspired by Section~\ref{Sec3}, a group draw with one set-based prohibited triplet list is considered: some teams are of type $S$, and at most two teams from type $S$ can be assigned to the same group.
There are three pots with $n$ teams each; pot A contains $i$ teams of type $S$, pot B contains $j$ teams of type $S$, and pot C contains $k$ teams of type $S$. Since all orders of the pots are examined, $i \geq j \geq k \geq 1$ can be assumed without loss of generality.

Under the assumptions above, six different sets of teams exist. Teams of type $S$ in pots A, B, C are collected in sets $S1$, $S2$, and $S3$, respectively, whereas teams of type not $S$ (which are not affected by the draw constraint) in pots A, B, C form sets $T1$, $T2$, and $T3$, respectively. Therefore, $\lvert S1 \rvert = i$, $\lvert S2 \rvert = j$, and $\lvert S3 \rvert = k$, as well as $\lvert T1 \rvert = n-i$, $\lvert T2 \rvert = n-j$, and $\lvert T3 \rvert = n-k$.

\begin{table}[t!]
  \centering
  \caption{Assignment probabilities in the theoretical model of Section~\ref{Sec4}}
  \label{Table1}
    \rowcolors{1}{gray!20}{}
    \begin{tabularx}{0.9\textwidth}{l CCCC c} \toprule 
    Event & Probability \\ \bottomrule
    Any team in $S1$ plays against any team in $S1$ & 0 \\
    Any team in $S1$ plays against any team in $S2$ & $p_{12}$ \\
    Any team in $S1$ plays against any team in $S3$ & $p_{13}$ \\
    Any team in $S2$ plays against any team in $S2$ & 0 \\
    Any team in $S2$ plays against any team in $S3$ & $p_{13}$ \\
    Any team in $S3$ plays against any team in $S3$ & 0 \\ \hline
    Any team in $S1$ plays against any team in $T1$ & 0 \\
    Any team in $S1$ plays against any team in $T2$ & $\left( 1 - jp_{12} \right)/(n-j)$ \\
    Any team in $S1$ plays against any team in $T3$ & $\left( 1 - kp_{13} \right)/(n-k)$ \\ \hline
    Any team in $S2$ plays against any team in $T1$ & $\left( 1 - ip_{12} \right)/(n-i)$ \\
    Any team in $S2$ plays against any team in $T2$ & 0 \\
    Any team in $S2$ plays against any team in $T3$ & $\left( 1 - kp_{23} \right)/(n-k)$ \\ \hline
    Any team in $S3$ plays against any team in $T1$ & $\left( 1 - ip_{13} \right)/(n-i)$ \\
    Any team in $S3$ plays against any team in $T2$ & $\left( 1 - jp_{23} \right)/(n-j)$ \\
    Any team in $S3$ plays against any team in $T3$ & 0 \\ \hline
    Any team in $T1$ plays against any team in $T1$ & 0 \\
    Any team in $T1$ plays against any team in $T2$ & $\left( 1 - i + ijp_{12} \right)/(n-i)(n-j)$ \\
    Any team in $T1$ plays against any team in $T3$ & $\left( 1 - i + ikp_{13} \right)/(n-i)(n-k)$ \\
    Any team in $T2$ plays against any team in $T2$ & 0 \\
    Any team in $T2$ plays against any team in $T3$ & $\left( 1 - j + jkp_{23} \right)/(n-j)(n-k)$ \\
    Any team in $T3$ plays against any team in $T3$ & 0 \\ \toprule
    \end{tabularx}
\end{table}

If a draw procedure does not distinguish between teams in the six sets of $S_1$, $S_2$, $S_3$, $T_1$, $T_2$, and $T_3$, then all assignment probabilities are determined by only three values, $p_{12}$ (the chance that any team in $S1$ is assigned to the same group as any team in $S2$), $p_{13}$ (the chance that any team in $S1$ is assigned to the same group as any team in $S3$), and $p_{23}$ (the chance that any team in $S2$ is assigned to the same group as any team in $S3$). They are collected in Table~\ref{Table1}.
The functions apply for both a uniform draw and the Skip mechanism with any draw order.
Naturally, $p_{12}$, $p_{13}$, and $p_{23}$ depend on the draw procedure, as we will see soon.

\subsection{General results for arbitrary numbers of teams} \label{Sec41}

Section~\ref{Sec411} determines the probabilities $p_{12}^U$, $p_{13}^U$, and $p_{23}^U$ under a uniform draw if $k=1$ (Proposition~\ref{Prop1}), which is sufficient to determine all assignment probabilities, see Table~\ref{Table1}.
Section~\ref{Sec412} assumes $j=k=1$ to derive the corresponding values under the Skip mechanism with its two distinct draw orders (Propositions~\ref{Prop2} and \ref{Prop3}). These preparations allow finding the least biased variant of the Skip mechanism according to two reasonable measures of aggregated distortions (Theorems~\ref{Theo1} and \ref{Theo2}).

\subsubsection{Probabilities under a uniform draw} \label{Sec411}

Under a uniform draw, all valid assignments have an equal chance to occur.

\begin{proposition} \label{Prop1}
Assume that the number of pots is three, the number of teams per pot is $n$, and at most two teams from set $S$ can be assigned to the same group. Furthermore, one pot contains $k=1$ team from set $S$.
Under a uniform draw, the assignment probabilities are:
\[
p_{12}^U = \frac{\left( n-i \right) + \left( i-1 \right) \cdot \left( n-j \right) / \left( n-1 \right)}{n^2 - ij},\qquad
p_{13}^U = \frac{n-j}{n^2 - ij},\qquad
p_{23}^U = \frac{n-i}{n^2 - ij}.
\]
\end{proposition}

\begin{proof}
Let us ignore the labels of the groups. Then the number of assignments in the absence of the draw constraint is $\left( n! \right)^2$. The draw constraint is violated if the only team in the set $S3$ is assigned to the same group as a team in $S1$ and another team in $S2$, which means $i \left( n-1 \right)! \cdot j \left( n-1 \right)!$ different assignments. Hence, the number of valid solutions is
\[
\left( n^2 - ij \right) \cdot \left[ \left( n-1 \right)! \right]^2.
\]

A team in $S1$ can play against the only team in $S3$ if their group does not contain any team in $S2$. This occurs in $\left( n-j \right) \left( n-1 \right)! \cdot \left( n-1 \right)!$ assignments, which provides the formula for $p_{13}^U$.

A team in $S2$ plays against the only team in $S3$ if their group does not contain any team in $S1$. Analogously to the previous case, this occurs in $\left( n-i \right) \left( n-1 \right)! \cdot \left( n-1 \right)!$ assignments, which provides the formula for $p_{23}^U$.

A team in $S1$ plays against a team in $S2$ if their group does not contain the only team in $S3$. To calculate the number of these assignments, two cases are distinguished:
\begin{itemize}
\item
The only team in $S3$ is assigned to a group that does not contain a team in $S1$ \\
This means $n-i$ different groups for the team in $S3$, thus, the corresponding number of admissible assignments is $\left( n-i \right) \left( n-1 \right)! \cdot \left( n-1 \right)!$.
\item
The only team in $S3$ is assigned to a group that does contain a team in $S1$ \\
This means $i-1$ different groups for the team in $S3$ (it cannot be assigned to the group where a team in $S1$ and a team in $S2$ play) such that its group should contain a team in $T2$ because of the draw constraint. The corresponding number of admissible assignments equals $\left( i-1 \right) \left( n-1 \right)! \cdot \left( n-j \right) \left( n-2 \right)!$.
\end{itemize}
To conclude,
\[
p_{12}^U = \frac{\left( n-i \right) \cdot \left( n-1 \right)! \cdot \left( n-1 \right)! + \left( i-1 \right) \cdot \left( n-1 \right)! \cdot \left( n-j \right) \cdot \left( n-2 \right)!}{\left( n^2 - ij \right) \left[ \left( n-1 \right)! \right]^2},
\]
which completes the proof as $\left( n-1 \right)! = (n-1) \cdot \left( n-2 \right)!$.
\end{proof}

\subsubsection{Probabilities under the Skip mechanism} \label{Sec412}

Now we turn to the case of the Skip mechanism. In contrast to Proposition~\ref{Prop1}, $j = 1$ is also assumed besides $k=1$, in order to reduce the complexity of the calculations.
Hence, only two distinct draw orders exist since pots B and C cannot be distinguished because both contain only one team of type $S$, and it does not count whether pot A is drawn first or second.
In particular, order A-B-C is equivalent to A-C-B, B-A-C, C-A-B, and order B-C-A is equivalent to C-B-A.


\begin{proposition} \label{Prop2}
Assume that the number of pots is three, the number of teams per pot is $n$, and at most two teams from set $S$ can be assigned to the same group. Furthermore, pot A contains $i \leq n-1$ teams from set $S$, while pots B and C contain $j = k = 1$ team from set $S$.
Under the Skip mechanism with the draw order A-B-C, the assignment probabilities are:
\[
p_{12}^{\mathit{ABC}} = \frac{1}{n},\qquad
p_{13}^{\mathit{ABC}} = \frac{n-1}{n} \cdot \frac{1}{n} + \frac{i-1}{n-1} \cdot \frac{1}{n} \cdot \frac{1}{n},\qquad
p_{23}^{\mathit{ABC}} = \frac{n-i}{n} \cdot \frac{1}{n}.
\]
\end{proposition}

\begin{proof}
A team in $S1$ plays against the only team in $S2$ if their draw positions coincide, which has a probability of $1/n$.

To derive the value of $p_{13}^{\mathit{ABC}}$, note that team in $S1$ plays against the only team in $S3$ in the following distinct cases:
\begin{itemize}
\item
If their draw positions coincide and the only team in $S2$ is not assigned to the same group, which has a probability of
\[
\frac{n-1}{n} \cdot \frac{1}{n}.
\]
\item
If the draw position of the team in $S1$ is one more than the draw position of another team in $S1$, of the only team in $S2$, as well as of the only team in $S3$ (when the latter skips a group due to the draw constraint), which has a probability of
\[
\frac{n-1}{n} \cdot \frac{i-1}{n-1} \cdot \frac{1}{n} \cdot \frac{1}{n}.
\]
\item
If the draw position of the team in $S1$ is $n-1$, and the draw position of another team in $S1$, of the only team in $S2$, as well as of the only team in $S3$ is $n$ (when the latter cannot be assigned to the last group due to the draw constraint, hence, it plays in the penultimate group---and the team in $T3$ that is drawn directly before the team in $S3$ is placed in the last group), which has a probability of
\[
\frac{1}{n} \cdot \frac{i-1}{n-1} \cdot \frac{1}{n} \cdot \frac{1}{n}.
\]
\end{itemize}

The only team in $S2$ plays against the only team in $S3$ if their draw positions coincide and no team in $S1$ has the same draw position, which has a probability of
\[
\frac{n-i}{n} \cdot \frac{1}{n}.
\]
\end{proof}

\begin{proposition} \label{Prop3}
Assume that the number of pots is three, the number of teams per pot is $n$, and at most two teams from set $S$ can be assigned to the same group. Furthermore, pot A contains $i \leq n-1$ teams from set $S$, while pots B and C contain $j = k = 1$ team from set $S$.
Under the Skip mechanism with the draw order B-C-A, the assignment probabilities are:
\[
p_{12}^{\mathit{BCA}} = \frac{n-1}{n} \cdot \frac{1}{n},\qquad
p_{13}^{\mathit{BCA}} = \frac{n-1}{n} \cdot \frac{1}{n},\qquad
p_{23}^{\mathit{BCA}} = \frac{1}{n}.
\]
\end{proposition}

\begin{proof}
A team in $S1$ plays against the only team in $S2$ ($S3$) if their draw positions coincide and the only team in $S3$ ($S2$) is not in the same draw position, which gives the expression for $p_{12}^{\mathit{BCA}}$ ($p_{13}^{\mathit{BCA}}$).

The only team in $S2$ plays against the only team in $S3$ if their draw positions coincide, which has a probability of $1/n$.
\end{proof}

Obviously, the number of cases to account for both in Propositions~\ref{Prop2} and \ref{Prop3} would be higher if $j \geq 2$. Furthermore, there would be three distinct draw orders instead of two.

\subsubsection{The optimal draw order of the Skip mechanism} \label{Sec313}

Now we can quantify the bias of the Skip mechanism with the draw orders A-B-C and B-C-A. However, some preparations are needed first. Denote the distortion of a draw mechanism for teams $s$ and $t$ compared to a uniform draw by $\tilde{p}_{st} = \lvert p_{st} - p_{st}^U \rvert$, where $p_{st}$ is the probability of assigning teams $s$ and $t$ to the same group under this particular draw mechanism, and $p_{st}^U$ is the probability under a uniform draw.

Since nine potentially different distortion values exist, both the mean (the sum of distortions for the $3n^2$ team pairs from different pots divided by $3n^2$) and the maximum of the biases are considered.

\begin{lemma} \label{Lemma3}
Assume that the number of pots is three, the number of teams per pot is $n$, and at most two teams from set $S$ can be assigned to the same group. Furthermore, pot A contains $i \leq n-1$ teams from set $S$, while pots B and C contain $j = k = 1$ team from set $S$.
The average distortion $\Delta$ of a draw mechanism is
\[
\Delta = \frac{4i \cdot \left( \tilde{p}_{12} + \tilde{p}_{13} \right) + 4 \tilde{p}_{23}}{3n^2}.
\]
\end{lemma}

\begin{proof}
There are $i$ teams in $S1$, and there is one team in $S2$. The distortion for these $i$ pairs of teams equals $\tilde{p}_{12}$.

There are $i$ teams in $S1$, and $n-1$ teams in $T2$. Based on Table~\ref{Table1}, the distortion for these $i \cdot (n-1)$ pairs of teams equals
\[
\left\lvert \frac{1- p_{12}}{n-1} - \frac{1- p_{12}^U}{n-1} \right\rvert = \left\lvert \frac{p_{12} - p_{12}^U}{n-1} \right\rvert = \frac{\tilde{p}_{12} }{n-1}.
\]

There are $n-i$ teams in $T1$, and there is one team in $S2$. Based on Table~\ref{Table1}, the distortion for these $n-i$ pairs of teams equals
\[
\left\lvert \frac{1- i \cdot p_{12}}{n-i} - \frac{1- i \cdot p_{12}^U}{n-i} \right\rvert = \left\lvert \frac{i \cdot \left( p_{12} - p_{12}^U \right)}{n-i} \right\rvert = \frac{i}{n-i} \cdot \tilde{p}_{12}.
\]

There are $n-i$ teams in $T1$, and $n-1$ teams in $T2$. Based on Table~\ref{Table1}, the distortion for these $(n-i) \cdot (n-1)$ pairs of teams equals
\[
\left\lvert \frac{1- i + i \cdot p_{12}}{(n-i) \cdot (n-1)} - \frac{1- i + i \cdot p_{12}^U}{(n-i) \cdot (n-1)} \right\rvert = \frac{i}{(n-i) \cdot (n-1)} \cdot \tilde{p}_{12}.
\]

Therefore, the sum of distortions for the sets $\left( S_1 \cup T_1 \right)$ and $\left( S_2 \cup T_2 \right)$ is $4i \cdot \tilde{p}_{12}$.

Analogous calculations uncover that the sum of distortions for the sets $\left( S_1 \cup T_1 \right)$ and $\left( S_3 \cup T_3 \right)$ is $4i \cdot \tilde{p}_{13}$.

Finally, there is one team in $S2$, and one team in $S3$. The distortion for this pair of teams equals $\tilde{p}_{23}$.

There is one team in $S2$, and there are $n-1$ teams in $T3$. Based on Table~\ref{Table1}, the distortion for these $n-1$ pairs of teams equals
\[
\left\lvert \frac{1- p_{23}}{n-1} - \frac{1- p_{23}^U}{n-1} \right\rvert = \left\lvert \frac{p_{23} - p_{23}^U}{n-1} \right\rvert = \frac{\tilde{p}_{23}}{n-1}.
\]

There are $n-1$ teams in $T2$, and there is one team in $S3$. Based on Table~\ref{Table1}, the distortion for these $n-1$ pairs of teams equals
\[
\left\lvert \frac{1- p_{23}}{n-1} - \frac{1- p_{23}^U}{n-1} \right\rvert = \left\lvert \frac{p_{23} - p_{23}^U}{n-1} \right\rvert = \frac{\tilde{p}_{23}}{n-1}.
\]

There are $n-1$ teams in $T2$, and $n-1$ teams in $T3$. Based on Table~\ref{Table1}, the distortion for these $(n-1) \cdot (n-1)$ pairs of teams equals
\[
\left\lvert \frac{p_{23}}{(n-1)^2} - \frac{p_{23}^U}{(n-1)^2} \right\rvert = \frac{\tilde{p}_{23}}{(n-1)^2}.
\]

Hence, the sum of distortions for the sets $\left( S_2 \cup T_2 \right)$ and $\left( S_3 \cup T_3 \right)$ is $4 \tilde{p}_{23}$.

The total number of team pairs from different pots is $3n^2$, which completes the proof.
\end{proof}

\begin{lemma} \label{Lemma4}
Assume that the number of pots is three, the number of teams per pot is $n$, and at most two teams from set $S$ can be assigned to the same group. Furthermore, pot A contains $i \leq n-1$ teams from set $S$, while pots B and C contain $j = k = 1$ team from set $S$.
The maximal distortion $\Lambda$ of a draw mechanism is
\[
\Lambda = \max \left\{ \tilde{p}_{12};\, \frac{i}{n-i} \cdot \tilde{p}_{12}; \, \tilde{p}_{13};\, \frac{i}{n-i} \cdot \tilde{p}_{13};\, \tilde{p}_{23} \right\}.
\]
\end{lemma}

\begin{proof}
See the calculations in the proof of Lemma~\ref{Lemma1}.
\end{proof}

Propositions~\ref{Prop1}--\ref{Prop3} provide the probabilities that should be substituted into Lemmata~\ref{Lemma3} and \ref{Lemma4} to compute the aggregated bias of the Skip mechanism with both draw orders.
As mentioned above, we will consider the average and the maximal distortions for all pairs of teams that have a nonzero probability of playing against each other.

\begin{proposition} \label{Prop4}
Assume that the number of pots is three, the number of teams per pot is $n$, and at most two teams from set $S$ can be assigned to the same group. Furthermore, pot A contains $i \leq n-1$ teams from set $S$, while pots B and C contain $j = k = 1$ team from set $S$.
The average distortion of the Skip mechanism with the draw order A-B-C is
\[
\Delta_{\mathit{ABC}} = \frac{4i(n-i) \left( n^2 + n - i - 1 \right)}{3n^4 (n-1) \left( n^2 - i \right)}.
\]
\end{proposition}

\begin{proof}
Based on Propositions~\ref{Prop1} and \ref{Prop2}:
\[
\tilde{p}_{12} = \left\lvert \frac{n-1}{n^2 - i} - \frac{1}{n} \right\lvert = \frac{n-i}{n \left( n^2 - i \right)};
\]
\[
\tilde{p}_{13} = \left\lvert \frac{n-1}{n^2 - i} - \frac{n-1}{n^2} - \frac{i-1}{n^2 (n-1)} \right\lvert = \frac{\left( n-i \right)^2}{n^2 (n-1) \left( n^2 - i \right)};
\]
\[
\tilde{p}_{23} = \left\lvert \frac{n-i}{n^2 - i} - \frac{n-i}{n^2} \right\lvert = \frac{i(n-i)}{n^2 \left( n^2 - i \right)}.
\]
Hence, Lemma~\ref{Lemma3} implies
\[
\Delta_{\mathit{ABC}} = \frac{4i (n-1) n (n-i) + 4i (n-i)^2 + 4i (n-1) (n-i)}{3n^4 (n-1) \left( n^2 - i \right)},
\]
which completes the proof after some simplifications.
\end{proof}

\begin{proposition} \label{Prop5}
Assume that the number of pots is three, the number of teams per pot is $n$, and at most two teams from set $S$ can be assigned to the same group. Furthermore, pot A contains $i \leq n-1$ teams from set $S$, while pots B and C contain $j = k = 1$ team from set $S$.
The maximal distortion of the Skip mechanism with the draw order A-B-C is
\[
\Lambda_{\mathit{ABC}} = \max \left\{ 1;\, \frac{i}{n-i} \right\} \cdot \frac{n-i}{n \left( n^2 - i \right)}.
\]
\end{proposition}

\begin{proof}
Based on the proof of Proposition~\ref{Prop4}:
\[
\frac{\tilde{p}_{12}}{\tilde{p}_{13}} = \frac{n(n-1)}{n-i} > 1;
\]
\[
\frac{\tilde{p}_{12}}{\tilde{p}_{23}} = \frac{n}{i} > 1.
\]
Thus, the first two terms in Lemma~\ref{Lemma4} are guaranteed to be greater than the last three, and the value of $\tilde{p}_{12}$ is known from the proof of Proposition~\ref{Prop4}.
\end{proof}

\begin{proposition} \label{Prop6}
Assume that the number of pots is three, the number of teams per pot is $n$, and at most two teams from set $S$ can be assigned to the same group. Furthermore, pot A contains $i \leq n-1$ teams from set $S$, while pots B and C contain $j = k = 1$ team from set $S$.
The average distortion of the Skip mechanism with the draw order B-C-A is
\[
\Delta_{\mathit{BCA}} = \frac{4i (n-1) (n + 2i)}{3n^4 \left( n^2 - i \right)}.
\]
\end{proposition}

\begin{proof}
Based on Propositions~\ref{Prop1} and \ref{Prop3}:
\[
\tilde{p}_{12} = \left\lvert \frac{n-1}{n^2 - i} - \frac{n-1}{n^2} \right\lvert = \frac{i (n-1)}{n^2 \left( n^2 - i \right)};
\]
\[
\tilde{p}_{13} = \left\lvert \frac{n-1}{n^2 - i} - \frac{n-1}{n^2} \right\lvert = \frac{i (n-1)}{n^2 \left( n^2 - i \right)};
\]
\[
\tilde{p}_{23} = \left\lvert \frac{n-i}{n^2 - i} - \frac{1}{n} \right\lvert = \frac{i(n-1)}{n \left( n^2 - i \right)}.
\]
Hence, Lemma~\ref{Lemma3} implies
\[
\Delta_{\mathit{BCA}} = \frac{8i^2 (n-1) + 4in (n-1)}{3n^4 \left( n^2 - i \right)},
\]
which completes the proof.
\end{proof}

\begin{proposition} \label{Prop7}
Assume that the number of pots is three, the number of teams per pot is $n$, and at most two teams from set $S$ can be assigned to the same group. Furthermore, pot A contains $i \leq n-1$ teams from set $S$, while pots B and C contain $j = k = 1$ team from set $S$.
The maximal distortion of the Skip mechanism with the draw order B-C-A is
\[
\Lambda_{\mathit{BCA}} = \frac{i(n-1)}{n \left( n^2 - i \right)}.
\]
\end{proposition}

\begin{proof}
Based on the proof of Proposition~\ref{Prop6}:
\[
\frac{\tilde{p}_{12}}{\tilde{p}_{13}} = 1;
\]
\[
\frac{\tilde{p}_{23}}{\tilde{p}_{12}} = n > \max \left\{ 1;\, \frac{i}{n-i} \right\}.
\]
Thus, the last term in Lemma~\ref{Lemma4} is guaranteed to be greater than the first four, and the value of $\tilde{p}_{23}$ is known from the proof of Proposition~\ref{Prop6}.
\end{proof}

Now the bias of the Skip mechanism with the two distinct draw orders can be compared in our theoretical model according to both measures.

\begin{theorem} \label{Theo1}
Assume that the number of pots is three, the number of teams per pot is $n$, and at most two teams from set $S$ can be assigned to the same group. Furthermore, pot A contains $2 \leq i \leq n-1$ teams from set $S$, while pots B and C contain $j = k = 1$ team from set $S$.
The average distortion of the Skip mechanism with the draw order A-B-C is smaller than its average distortion with the draw order B-C-A.
\end{theorem}

\begin{proof}
According to Propositions~\ref{Prop4} and \ref{Prop6}:
\begin{equation} \label{eq_Delta}
\frac{\Delta_{\mathit{ABC}}}{\Delta_{\mathit{BCA}}} = \frac{(n-i) \left( n^2 + n - i - 1 \right)}{(n + 2i) \left( n^2 - 2n + 1 \right)} = \frac{n^3 + (1-i) n^2 - (2i + 1)n + i (i+1)}{n^3 + (2i - 2 )n^2 + (1 - 4i)n + 2i}.
\end{equation}
The numerator of \eqref{eq_Delta} is smaller than $a = n^3 - n^2 - (i+1)n + i$ since $i^2 < in$ and $i \geq 2$. The denominator of \eqref{eq_Delta} is greater than $b = n^3 - n^2 + (2in - n - 4i + 1)n + i$. Furthermore, $b - a = (2i-1)n - 4i + 1 + (i+1) > 0$ because $n > (3i-2) / (2i-1)$ if $n \geq 2$. This completes the proof as the value of \eqref{eq_Delta} remains below one.
\end{proof}

\begin{theorem} \label{Theo2}
Assume that the number of pots is three, the number of teams per pot is $n$, and at most two teams from set $S$ can be assigned to the same group. Furthermore, pot A contains $2 \leq i \leq n-1$ teams from set $S$, while pots B and C contain $j = k = 1$ team from set $S$.
The maximal distortion of the Skip mechanism with the draw order A-B-C is smaller than its maximal distortion with the draw order B-C-A.
\end{theorem}

\begin{proof}
According to Propositions~\ref{Prop5} and \ref{Prop7}:
\begin{equation} \label{eq_Lambda}
\frac{\Lambda_{\mathit{ABC}}}{\Lambda_{\mathit{BCA}}} =  \max \left\{ 1;\, \frac{i}{n-i} \right\} \cdot \frac{n-i}{i(n-1)}.
\end{equation}
\eqref{eq_Lambda} is smaller than $1/i$ if $i/(n-i) \leq 1$ and equal to $1/(n-1)$ if $i/(n-i) > 1$.
\end{proof}

\begin{figure}[t!]
\centering

\begin{tikzpicture}
\begin{axis}[
name = axis1,
width = 0.48\textwidth, 
height = 0.4\textwidth,
title = {$n=3$ teams per pot},
title style = {font=\small},
xmajorgrids,
ymajorgrids,
xlabel = {Value of $i$},
xlabel style = {align=center, font=\small},
scaled x ticks = false,
xtick = data,
ylabel = {Distortion in \% (log scale)},
ylabel style = {align=center, font=\small},
ytick style = {draw = none},
ymode = log,
log ticks with fixed point,
ymin = 0.008,
ymax = 50,
]
\addplot [blue, only marks, mark = square, thick] coordinates{
(2,2.11640211640212)
};
\addplot [red, only marks, mark = star, thick] coordinates{
(2,9.52380952380952)
};
\addplot [ForestGreen, only marks, mark = triangle, very thick] coordinates{
(2,6.58436213991769)
};
\addplot [black, only marks, mark = pentagon, thick] coordinates{
(2,19.047619047619)
};
\end{axis}

\begin{axis}[
at = {(axis1.south east)},
xshift = 0.12\textwidth,
width = 0.48\textwidth, 
height = 0.4\textwidth,
title = {$n=4$ teams per pot},
title style = {font=\small},
xmajorgrids,
ymajorgrids,
xlabel = {Value of $i$},
xlabel style = {align=center, font=\small},
enlarge x limits = 0.5,
scaled x ticks = false,
xtick = data,
ylabel = {Distortion in \% (log scale)},
ylabel style = {align=center, font=\small},
ytick style = {draw = none},
ymode = log,
log ticks with fixed point,
ymin = 0.008,
ymax = 50,
]
\addplot [blue, only marks, mark = square, thick] coordinates{
(2,0.843253968253968)
(3,0.641025641025641)
};
\addplot [red, only marks, mark = star, thick] coordinates{
(2,3.57142857142857)
(3,5.76923076923077)
};
\addplot [ForestGreen, only marks, mark = triangle, very thick] coordinates{
(2,1.78571428571429)
(3,3.60576923076923)
};
\addplot [black, only marks, mark = pentagon, thick] coordinates{
(2,10.7142857142857)
(3,17.3076923076923)
};
\end{axis}
\end{tikzpicture}

\vspace{0.5cm}
\begin{tikzpicture}
\begin{axis}[
name = axis3,
width = 0.48\textwidth, 
height = 0.4\textwidth,
title = {$n=5$ teams per pot},
title style = {font=\small},
xmajorgrids,
ymajorgrids,
xlabel = {Value of $i$},
xlabel style = {align=center, font=\small},
enlarge x limits = 0.2,
scaled x ticks = false,
xtick = data,
ylabel = {Distortion in \% (log scale)},
ylabel style = {align=center, font=\small},
ytick style = {draw = none},
ymode = log,
log ticks with fixed point,
ymin = 0.008,
ymax = 50,
]
\addplot [blue, only marks, mark = square, thick] coordinates{
(2,0.375652173913044)
(3,0.378181818181818)
(4,0.253968253968254)
};
\addplot [red, only marks, mark = star, thick] coordinates{
(2,2.60869565217391)
(3,2.72727272727273)
(4,3.80952380952381)
};
\addplot [ForestGreen, only marks, mark = triangle, very thick] coordinates{
(2,0.667826086956522)
(3,1.28)
(4,2.11301587301587)
};
\addplot [black, only marks, mark = pentagon, thick] coordinates{
(2,6.95652173913043)
(3,10.9090909090909)
(4,15.2380952380952)
};
\end{axis}

\begin{axis}[
at = {(axis3.south east)},
xshift = 0.12\textwidth,
width = 0.48\textwidth, 
height = 0.4\textwidth,
title = {$n=6$ teams per pot},
title style = {font=\small},
xmajorgrids,
ymajorgrids,
xlabel = {Value of $i$},
xlabel style = {align=center, font=\small},
enlarge x limits = 0.2,
scaled x ticks = false,
xtick = data,
ylabel = {Distortion in \% (log scale)},
ylabel style = {align=center, font=\small},
ytick style = {draw = none},
ymode = log,
log ticks with fixed point,
ymin = 0.008,
ymax = 50,
]
\addplot [blue, only marks, mark = square, thick] coordinates{
(2,0.18881626724764)
(3,0.21324354657688)
(4,0.190329218106996)
(5,0.1194743130227)
};
\addplot [red, only marks, mark = star, thick] coordinates{
(2,1.96078431372549)
(3,1.51515151515152)
(4,2.08333333333333)
(5,2.68817204301075)
};
\addplot [ForestGreen, only marks, mark = triangle, very thick] coordinates{
(2,0.302590171871218)
(3,0.561167227833894)
(4,0.900205761316872)
(5,1.32749236691889)
};
\addplot [black, only marks, mark = pentagon, thick] coordinates{
(2,4.90196078431373)
(3,7.57575757575758)
(4,10.4166666666667)
(5,13.4408602150538)
};
\end{axis}
\end{tikzpicture}

\vspace{0.5cm}
\begin{tikzpicture}
\begin{axis}[
name = axis5,
width = 0.48\textwidth, 
height = 0.4\textwidth,
title = {$n=8$ teams per pot},
title style = {font=\small},
xmajorgrids,
ymajorgrids,
xlabel = {Value of $i$},
xlabel style = {align=center, font=\small},
enlarge x limits = 0.1,
scaled x ticks = false,
xtick = data,
ylabel = {Distortion in \% (log scale)},
ylabel style = {align=center, font=\small},
ytick style = {draw = none},
ymode = log,
log ticks with fixed point,
ymin = 0.008,
ymax = 50,
]
\addplot [blue, only marks, mark = square, thick] coordinates{
(2,0.0621039746543779)
(3,0.077759074941452)
(4,0.0830853174603175)
(5,0.0780304176755448)
(6,0.0625384852216749)
(7,0.0365497076023392)
};
\addplot [red, only marks, mark = star, thick] coordinates{
(2,1.20967741935484)
(3,1.02459016393443)
(4,0.833333333333333)
(5,1.05932203389831)
(6,1.29310344827586)
(7,1.53508771929825)
};
\addplot [ForestGreen, only marks, mark = triangle, very thick] coordinates{
(2,0.0882056451612903)
(3,0.156890368852459)
(4,0.243055555555556)
(5,0.347590042372881)
(6,0.471443965517241)
(7,0.615634137426901)
};
\addplot [black, only marks, mark = pentagon, thick] coordinates{
(2,2.82258064516129)
(3,4.30327868852459)
(4,5.83333333333333)
(5,7.41525423728813)
(6,9.05172413793103)
(7,10.7456140350877)
};
\end{axis}

\begin{axis}[
at = {(axis5.south east)},
xshift = 0.12\textwidth,
width = 0.48\textwidth, 
height = 0.4\textwidth,
title = {$n=10$ teams per pot},
title style = {font=\small},
xmajorgrids,
ymajorgrids,
xlabel = {Value of $i$},
xlabel style = {align=center, font=\small},
enlarge x limits = 0.1,
scaled x ticks = false,
xtick = data,
ylabel = {Distortion in \% (log scale)},
ylabel style = {align=center, font=\small},
ytick style = {draw = none},
ymode = log,
log ticks with fixed point,
ymin = 0.008,
ymax = 50,
legend style = {font=\small,at={(-1.3,-0.25)},anchor=north west,legend columns=4},
legend entries = {Average $\Delta_{\mathit{ABC}} \qquad$, Maximal $\Lambda_{\mathit{ABC}} \qquad$, Average $\Delta_{\mathit{BCA}} \qquad$, Maximal $\Lambda_{\mathit{BCA}}$},
]
\addplot [blue, only marks, mark = square, thick] coordinates{
(2,0.025880574452003)
(3,0.0339977090492554)
(4,0.0388888888888889)
(5,0.0405458089668616)
(6,0.0389598108747045)
(7,0.0341218637992832)
(8,0.0260225442834138)
(9,0.0146520146520147)
};
\addplot [red, only marks, mark = star, thick] coordinates{
(2,0.816326530612245)
(3,0.721649484536082)
(4,0.625)
(5,0.526315789473684)
(6,0.638297872340426)
(7,0.752688172043011)
(8,0.869565217391304)
(9,0.989010989010989)
};
\addplot [ForestGreen, only marks, mark = triangle, very thick] coordinates{
(2,0.0342857142857143)
(3,0.0593814432989691)
(4,0.09)
(5,0.126315789473684)
(6,0.168510638297872)
(7,0.216774193548387)
(8,0.271304347826087)
(9,0.332307692307692)
};
\addplot [black, only marks, mark = pentagon, thick] coordinates{
(2,1.83673469387755)
(3,2.78350515463918)
(4,3.75)
(5,4.73684210526316)
(6,5.74468085106383)
(7,6.7741935483871)
(8,7.82608695652174)
(9,8.9010989010989)
};
\end{axis}
\end{tikzpicture}

\caption{Distortion of the Skip mechanism, theoretical model with $j=k=1$}
\label{Fig1}

\end{figure}


Figure~\ref{Fig1} visualises the bias of the Skip mechanism with the two distinct draw orders according to the two aggregated measures for selected values of $n$.
As expected, a higher number of teams ($n$) reduces the bias if the number of teams affected by the draw constraint ($i$) remains fixed. While both the average and maximal distortions are increasing as a function of $i$ under the inferior draw order B-C-A, having more restrictions is not always unfavourable under the optimal draw order A-B-C. However, the maximal bias is relatively insensitive to $i$, and exceeds 0.5\% even if there are $n=10$ teams in each pot.

\begin{figure}
\centering

\begin{subfigure}{0.495\textwidth}
\centering
\caption{Reduction of absolute distortion $\Delta$}
\label{Fig2a}

\resizebox{\textwidth}{!}{
\begin{tikzpicture}
\node at (3,1) {\large{3}};
\node at (6,1) {\large{6}};
\node at (9,1) {\large{9}};
\node at (12,1) {\large{12}};
\node at (7.5,0) {\large{Number of teams in each pot ($n$)}};
\node at (2,2) {\large{2}};
\node at (2,5) {\large{5}};
\node at (2,8) {\large{8}};
\node at (2,11) {\large{11}};
\node[rotate=90] at (1,6.5) {\large{Teams in Pot 1 with a draw constraint ($i$)}};

\fill[cyan] (3,2) circle (0.339285714285714);
\fill[cyan] (4,2) circle (0.263888888888889);
\fill[cyan] (4,3) circle (0.411111111111111);
\fill[cyan] (5,2) circle (0.21875);
\fill[cyan] (5,3) circle (0.352272727272727);
\fill[cyan] (5,4) circle (0.439903846153846);
\fill[cyan] (6,2) circle (0.188);
\fill[cyan] (6,3) circle (0.31);
\fill[cyan] (6,4) circle (0.394285714285714);
\fill[cyan] (6,5) circle (0.455);
\fill[cyan] (7,2) circle (0.16540404040404);
\fill[cyan] (7,3) circle (0.277777777777778);
\fill[cyan] (7,4) circle (0.358333333333333);
\fill[cyan] (7,5) circle (0.418300653594771);
\fill[cyan] (7,6) circle (0.464181286549708);
\fill[cyan] (8,2) circle (0.147959183673469);
\fill[cyan] (8,3) circle (0.252186588921283);
\fill[cyan] (8,4) circle (0.329081632653061);
\fill[cyan] (8,5) circle (0.387755102040816);
\fill[cyan] (8,6) circle (0.433673469387755);
\fill[cyan] (8,7) circle (0.470315398886827);
\fill[cyan] (9,2) circle (0.134014423076923);
\fill[cyan] (9,3) circle (0.23125);
\fill[cyan] (9,4) circle (0.3046875);
\fill[cyan] (9,5) circle (0.361842105263158);
\fill[cyan] (9,6) circle (0.407366071428571);
\fill[cyan] (9,7) circle (0.44429347826087);
\fill[cyan] (9,8) circle (0.4746875);
\fill[cyan] (10,2) circle (0.122574955908289);
\fill[cyan] (10,3) circle (0.213734567901235);
\fill[cyan] (10,4) circle (0.283950617283951);
\fill[cyan] (10,5) circle (0.339506172839506);
\fill[cyan] (10,6) circle (0.384399551066218);
\fill[cyan] (10,7) circle (0.421296296296296);
\fill[cyan] (10,8) circle (0.452041785375119);
\fill[cyan] (10,9) circle (0.477954144620811);
\fill[cyan] (11,2) circle (0.113);
\fill[cyan] (11,3) circle (0.198823529411765);
\fill[cyan] (11,4) circle (0.266052631578947);
\fill[cyan] (11,5) circle (0.32);
\fill[cyan] (11,6) circle (0.364130434782609);
\fill[cyan] (11,7) circle (0.4008);
\fill[cyan] (11,8) circle (0.431666666666667);
\fill[cyan] (11,9) circle (0.457931034482759);
\fill[cyan] (11,10) circle (0.480483870967742);
\fill[cyan] (12,2) circle (0.104855371900826);
\fill[cyan] (12,3) circle (0.18595041322314);
\fill[cyan] (12,4) circle (0.250413223140496);
\fill[cyan] (12,5) circle (0.302779864763336);
\fill[cyan] (12,6) circle (0.346074380165289);
\fill[cyan] (12,7) circle (0.382390336935791);
\fill[cyan] (12,8) circle (0.413223140495868);
\fill[cyan] (12,9) circle (0.439669421487603);
\fill[cyan] (12,10) circle (0.462551652892562);
\fill[cyan] (12,11) circle (0.482498784637822);
\end{tikzpicture}
}
\end{subfigure}
\begin{subfigure}{0.495\textwidth}
\centering
\caption{Reduction of maximal distortion $\Lambda$}
\label{Fig2b}

\resizebox{\textwidth}{!}{
\begin{tikzpicture}
\node at (3,1) {\large{3}};
\node at (6,1) {\large{6}};
\node at (9,1) {\large{9}};
\node at (12,1) {\large{12}};
\node at (7.5,0) {\large{Number of teams in each pot ($n$)}};
\node at (2,2) {\large{2}};
\node at (2,5) {\large{5}};
\node at (2,8) {\large{8}};
\node at (2,11) {\large{11}};
\node[rotate=90] at (1,6.5) {\large{Teams in Pot 1 with a draw constraint ($i$)}};

\fill[ForestGreen] (3,2) circle (0.25);
\fill[ForestGreen] (4,2) circle (0.333333333333333);
\fill[ForestGreen] (4,3) circle (0.333333333333333);
\fill[ForestGreen] (5,2) circle (0.3125);
\fill[ForestGreen] (5,3) circle (0.375);
\fill[ForestGreen] (5,4) circle (0.375);
\fill[ForestGreen] (6,2) circle (0.3);
\fill[ForestGreen] (6,3) circle (0.4);
\fill[ForestGreen] (6,4) circle (0.4);
\fill[ForestGreen] (6,5) circle (0.4);
\fill[ForestGreen] (7,2) circle (0.291666666666667);
\fill[ForestGreen] (7,3) circle (0.388888888888889);
\fill[ForestGreen] (7,4) circle (0.416666666666667);
\fill[ForestGreen] (7,5) circle (0.416666666666667);
\fill[ForestGreen] (7,6) circle (0.416666666666667);
\fill[ForestGreen] (8,2) circle (0.285714285714286);
\fill[ForestGreen] (8,3) circle (0.380952380952381);
\fill[ForestGreen] (8,4) circle (0.428571428571429);
\fill[ForestGreen] (8,5) circle (0.428571428571429);
\fill[ForestGreen] (8,6) circle (0.428571428571429);
\fill[ForestGreen] (8,7) circle (0.428571428571429);
\fill[ForestGreen] (9,2) circle (0.28125);
\fill[ForestGreen] (9,3) circle (0.375);
\fill[ForestGreen] (9,4) circle (0.421875);
\fill[ForestGreen] (9,5) circle (0.4375);
\fill[ForestGreen] (9,6) circle (0.4375);
\fill[ForestGreen] (9,7) circle (0.4375);
\fill[ForestGreen] (9,8) circle (0.4375);
\fill[ForestGreen] (10,2) circle (0.277777777777778);
\fill[ForestGreen] (10,3) circle (0.37037037037037);
\fill[ForestGreen] (10,4) circle (0.416666666666667);
\fill[ForestGreen] (10,5) circle (0.444444444444444);
\fill[ForestGreen] (10,6) circle (0.444444444444444);
\fill[ForestGreen] (10,7) circle (0.444444444444444);
\fill[ForestGreen] (10,8) circle (0.444444444444444);
\fill[ForestGreen] (10,9) circle (0.444444444444444);
\fill[ForestGreen] (11,2) circle (0.275);
\fill[ForestGreen] (11,3) circle (0.366666666666667);
\fill[ForestGreen] (11,4) circle (0.4125);
\fill[ForestGreen] (11,5) circle (0.44);
\fill[ForestGreen] (11,6) circle (0.45);
\fill[ForestGreen] (11,7) circle (0.45);
\fill[ForestGreen] (11,8) circle (0.45);
\fill[ForestGreen] (11,9) circle (0.45);
\fill[ForestGreen] (11,10) circle (0.45);
\fill[ForestGreen] (12,2) circle (0.272727272727273);
\fill[ForestGreen] (12,3) circle (0.363636363636364);
\fill[ForestGreen] (12,4) circle (0.409090909090909);
\fill[ForestGreen] (12,5) circle (0.436363636363636);
\fill[ForestGreen] (12,6) circle (0.454545454545455);
\fill[ForestGreen] (12,7) circle (0.454545454545455);
\fill[ForestGreen] (12,8) circle (0.454545454545455);
\fill[ForestGreen] (12,9) circle (0.454545454545455);
\fill[ForestGreen] (12,10) circle (0.454545454545455);
\fill[ForestGreen] (12,11) circle (0.454545454545455);
\end{tikzpicture}
}
\end{subfigure}

\captionsetup{justification=centerfirst}
\caption{The benefit of the optimal draw order, theoretical model with $j=k=1$ \\ \vspace{0.2cm}
\footnotesize{\emph{Note}: The size of the dots is proportional to the improvement of unfairness.}}
\label{Fig2}
\end{figure}


Figure~\ref{Fig2} shows the gain from using the optimal draw order A-B-C. The bias is more than halved with respect to maximal distortion (Figure~\ref{Fig2b}), and its relative decrease reaches $(n-2)/(n-1)$ for high values of $i$ as demonstrated by Theorem~\ref{Theo2}. The advantage of the draw order A-B-C is far from being marginal for average distortion, too, and is greater than the relative decrease in the maximal distortion if $i$ is close to $n$, that is, the draw constraint applies to almost all teams in pot A. For instance, the average bias $\Delta$ is reduced by more than 92.5\% if $n=12$ and $10 \leq i \leq 11$.

\subsection{All possible cases up to five teams in each pot} \label{Sec42}

The results presented in Section~\ref{Sec41} apply to an arbitrary number of teams, but the draw constraint affects only one team in pots 2 and 3 each. Since extending the expressions of $p_{12}$, $p_{13}$, and $p_{23}$ in Propositions~\ref{Prop2} and \ref{Prop3} to $j > 1$ and $k > 1$ would be complicated, we have decided for numerical calculations up to $3n = 15$ teams. Note that the number of orders in which the teams can be drawn is $(n!)^3$ for each draw order, which is 1.728 million if $n=5$, but exceeds 37 million if $n=6$.

Due to the set-based prohibited triplet list, only three different draw orders exist as the order of the first two pots does not count, namely, draw order A-B-C coincides with B-A-C, draw order A-C-B coincides with C-A-B, and draw order B-C-A coincides with C-B-A.
In addition, the number of different draw orders is two if $i=j$ or $j=k$, and only one if $i=j=k$.

\begin{table}[t!]
  \centering
  \caption{Distortion of the Skip mechanism, theoretical model with $n \leq 5$}
  \label{Table2}
     \rowcolors{1}{}{gray!20}
\centerline{
\begin{threeparttable}
    \begin{tabularx}{1.15\textwidth}{ccccc CCr CCr} \toprule \hiderowcolors
    \multicolumn{4}{c}{\textbf{Parameters}} & Proportion & \multicolumn{3}{c}{\textbf{Average distortion $\Delta$}} &\multicolumn{3}{c}{\textbf{Maximal distortion $\Lambda$}} \\
    \multirow{2}[0]{*}{$n$} & \multirow{2}[0]{*}{$i$} & \multirow{2}[0]{*}{$j$} & \multirow{2}[0]{*}{$k$} & of valid & \multicolumn{2}{c}{Draw order} & \multicolumn{1}{r}{\multirow{2}[0]{*}{Reduction}} & \multicolumn{2}{c}{Draw order} & \multicolumn{1}{r}{\multirow{2}[0]{*}{Reduction}} \\
          &       &       &       & assignments & A-B-C & B-C-A &       & A-B-C & B-C-A &  \\ \bottomrule \showrowcolors
    3     & 1     & 1     & 1     & 88.89\% & 2.058\% & 2.058\% & ---   & \textcolor{gray!20}{0}8.333\% & \textcolor{gray!20}{0}8.333\% & --- \\
    3     & 2     & 1     & 1     & 77.78\% & 2.116\% & 6.584\% & 67.86\% & \textcolor{white}{0}9.524\% & 19.048\% & 50.00\% \\
    3     & 2     & 2     & 1     & 55.56\% & 3.951\% & 7.901\% & 50.00\% & 13.333\% & 26.667\% & 50.00\% \\
    3     & 2     & 2     & 2     & 22.22\% & 0     & 0     & 0     & 0     & 0     & 0 \\ \hline
    4     & 1     & 1     & 1     & 93.75\% & 0.625\% & 0.625\% & ---   & \textcolor{gray!20}{0}5.000\% & \textcolor{gray!20}{0}5.000\% & --- \\
    4     & 2     & 1     & 1     & 87.5\% & 0.843\% & 1.786\% & 52.78\% & \textcolor{white}{0}3.571\% & 10.714\% & 66.67\% \\
    4     & 2     & 2     & 1     & 75\%  & 1.389\% & 2.546\% & 45.45\% & \textcolor{gray!20}{0}2.778\% & \textcolor{gray!20}{0}8.333\% & 66.67\% \\
    4     & 2     & 2     & 2     & 52.78\% & 4.288\% & 4.288\% & ---   & \textcolor{white}{0}6.579\% & \textcolor{white}{0}6.579\% & --- \\
    4     & 3     & 1     & 1     & 81.25\% & 0.641\% & 3.606\% & 82.22\% & \textcolor{gray!20}{0}5.769\% & 17.308\% & 66.67\% \\
    4     & 3     & 2     & 1     & 62.5\% & 1.250\% & 5.625\% & 77.78\% & \textcolor{white}{0}5.000\% & 22.500\% & 77.78\% \\
    4     & 3     & 2     & 2     & 33.33\% & 4.167\% & 5.556\% & 25.00\% & 12.500\% & \textcolor{gray!20}{0}8.333\% & $-$50.00\% \\
    4     & 3     & 3     & 1     & 43.75\% & 1.786\% & 5.357\% & 66.67\% & 10.714\% & 32.143\% & 66.67\% \\
    4     & 3     & 3     & 2     & 12.5\% & 0     & 0     & 0     & 0     & 0     & 0 \\ \hline
    5     & 1     & 1     & 1     & 96\%  & 0.249\% & 0.249\% & ---   & \textcolor{white}{0}3.333\% & \textcolor{white}{0}3.333\% & --- \\
    5     & 2     & 1     & 1     & 92\%  & 0.376\% & 0.668\% & 43.75\% & \textcolor{gray!20}{0}2.609\% & \textcolor{gray!20}{0}6.957\% & 62.50\% \\
    5     & 2     & 2     & 1     & 84\%  & 0.625\% & 1.036\% & 39.71\% & \textcolor{white}{0}2.143\% & \textcolor{white}{0}5.714\% & 62.50\% \\
    5     & 2     & 2     & 2     & 69\%  & 1.753\% & 1.753\% & ---   & \textcolor{gray!20}{0}4.783\% & \textcolor{gray!20}{0}4.783\% & --- \\
    5     & 3     & 1     & 1     & 88\%  & 0.378\% & 1.280\% & 70.45\% & \textcolor{white}{0}2.727\% & 10.909\% & 75.00\% \\
    5     & 3     & 2     & 1     & 76\%  & 0.688\% & 2.072\% & 66.80\% & \textcolor{gray!20}{0}2.368\% & \textcolor{gray!20}{0}9.474\% & 75.00\% \\
    5     & 3     & 2     & 2     & 55\%  & 1.988\% & 3.578\% & 44.44\% & \textcolor{white}{0}5.455\% & \textcolor{white}{0}8.182\% & 33.33\% \\
    5     & 3     & 3     & 1     & 64\%  & 0.867\% & 2.280\% & 61.99\% & \textcolor{gray!20}{0}2.813\% & 11.250\% & 75.00\% \\
    5     & 3     & 3     & 2     & 37\%  & 2.566\% & 4.216\% & 39.15\% & \textcolor{white}{0}6.486\% & \textcolor{white}{0}9.730\% & 33.33\% \\
    5     & 3     & 3     & 3     & 18\%  & 3.911\% & 3.911\% & ---   & \textcolor{gray!20}{0}9.167\% & \textcolor{gray!20}{0}9.167\% & --- \\
    5     & 4     & 1     & 1     & 84\%  & 0.254\% & 2.113\% & 87.98\% & \textcolor{white}{0}3.810\% & 15.238\% & 75.00\% \\
    5     & 4     & 2     & 1     & 68\%  & 0.502\% & 3.614\% & 86.11\% & \textcolor{gray!20}{0}3.529\% & 22.588\% & 84.38\% \\
    5     & 4     & 2     & 2     & 42\%  & 1.524\% & 4.861\% & 68.65\% & \textcolor{white}{0}8.571\% & 12.429\% & 31.03\% \\
    5     & 4     & 3     & 1     & 52\%  & 0.738\% & 4.332\% & 82.95\% & \textcolor{gray!20}{0}4.615\% & 29.538\% & 84.38\% \\
    5     & 4     & 3     & 2     & 22\%  & 2.327\% & 3.782\% & 38.46\% & 10.909\% & 10.636\% & $-$2.56\% \\
    5     & 4     & 3     & 3     & \textcolor{gray!20}{0}6\%   & 0     & 0     & 0     & 0     & 0     & 0 \\
    5     & 4     & 4     & 1     & 36\%  & 0.948\% & 3.793\% & 75.00\% & \textcolor{white}{0}8.889\% & 35.556\% & 75.00\% \\
    5     & 4     & 4     & 2     & \textcolor{gray!20}{0}8\%   & 0     & 0     & 0     & 0     & 0     & 0 \\ \toprule
    \end{tabularx}
\begin{tablenotes} \footnotesize
\item
\emph{Notes:} The column Reduction shows the relative advantage of the draw order A-B-C over the draw order B-C-A. \\
See Figure~\ref{Fig3} for the draw order A-C-B.
\end{tablenotes}
\end{threeparttable} 
}   
\end{table}

Table~\ref{Table2} summarises the aggregated distortions for all cases if $n \leq 5$. $i \leq n-1$ is assumed since $i=n$ would strongly simplify the draw constraint and ease the problem faced by the Skip mechanism. For $n=4$, there is no valid assignment if $i = j = k = 3$ because nine teams cannot be assigned to four groups such that no group contains more than two teams. Analogously, $n=5$ does not allow for $i = j = 4$ and $k \geq 3$.

Four sets of parameters guarantee that the Skip mechanism (with any draw order) has uniform distribution as the constraint severely restricts the number of solutions. Average distortion with the draw order A-B-C is always smaller than with the draw order B-C-A, except if all draw orders coincide due to the commutability of the three pots. Its advantage is non-negligible, the relative gain is at least 25\% and can exceed 80\%. Both findings are in line with Theorem~\ref{Theo1}, although its assumptions do not hold.

On the other hand, the average (maximal) distortion $\Delta$ ($\Lambda$) is certainly above 2\% (8\%) for some parameters if $n=3$. Analogously, there is a scenario with $\Delta$ ($\Lambda$) higher than 3.9\% (8\%) for both $n=4$ and $n=5$. Consequently, a substantial scope remains to mitigate the distortions of the Skip mechanism by a novel randomisation procedure, and the performance of other draw mechanisms is worth investigating in this relatively simple setting.

Last but not least, Theorem~\ref{Theo2} cannot be generalised to $j \geq k \geq 2$. In particular, maximal distortion $\Lambda$ under the draw order B-C-A is substantially smaller than under the draw order A-B-C if $n=4$, $i=3$, and $j=k=2$. However, the average bias $\Delta$ is more favourable under the draw order A-B-C, and the gain in maximal distortion in the analogous case for $n=5$ (when the parameters are $i=4$, $j=3$, $k=2$) is marginal. Hence, these two anomalies may be caused only by the small number of teams.

\begin{figure}[t!]
\centering

\begin{subfigure}{\textwidth}
\centering

\caption{Average distortion $\Delta$}
\label{Fig3a}

\begin{tikzpicture}
\begin{axis}[
width = 1\textwidth, 
height = 0.6\textwidth,
xmajorgrids,
ymajorgrids,
xlabel = {Proportion of valid assignments},
xlabel style = {align=center, font=\small},
xmin = 0,
xmax = 1,
ylabel = {Distortion in \%},
ylabel style = {align=center, font=\small},
ymin = 0,
legend style = {font=\small,at={(0.02,-0.15)},anchor=north west,legend columns=3},
legend entries = {Order A-B-C$\qquad \qquad \quad \qquad \, \;$,Orders A-B-C and A-C-B$\qquad$,Order A-C-B,Orders A-C-B and B-C-A$\qquad$,Order B-C-A$\qquad \qquad \qquad \quad \;$,All orders$\quad \;$},
]
\addplot [blue, only marks, mark = square, thick] coordinates{
(0.555555555555556,3.95061728395062)
(0.75,1.38888888888889)
(0.625,1.25)
(0.4375,1.78571428571428)
(0.84,0.624761904761904)
(0.76,0.687719298245615)
(0.64,0.866666666666667)
(0.37,2.56576576576577)
(0.68,0.501960784313725)
(0.52,0.738461538461538)
(0.22,2.32727272727273)
(0.36,0.948148148148147)
};
\addplot [red, only marks, mark = star, thick] coordinates{
(0.777777777777778,2.11640211640212)
(0.875,0.843253968253969)
(0.8125,0.64102564102564)
(0.333333333333333,4.16666666666667)
(0.92,0.375652173913044)
(0.88,0.378181818181818)
(0.55,1.98787878787879)
(0.84,0.253968253968254)
(0.42,1.52380952380952)
};
\addplot [black, only marks, mark = pentagon, thick] coordinates{
(0.625,2.08333333333333)
(0.76,1.07789473684211)
(0.68,0.752941176470588)
(0.52,1.72307692307692)
(0.22,3.49090909090909)
};
\addplot [orange, only marks, mark = diamond, thick] coordinates{
(0.555555555555556,7.90123456790123)
(0.75,2.54629629629629)
(0.4375,5.35714285714286)
(0.84,1.03619047619048)
(0.64,2.28)
(0.37,4.21621621621622)
(0.36,3.7925925925926)
};
\addplot [ForestGreen, only marks, mark = triangle, thick] coordinates{
(0.777777777777778,6.5843621399177)
(0.875,1.78571428571428)
(0.8125,3.60576923076923)
(0.625,5.625)
(0.333333333333333,5.55555555555555)
(0.92,0.667826086956521)
(0.88,1.28)
(0.76,2.07157894736842)
(0.55,3.57818181818181)
(0.84,2.11301587301587)
(0.68,3.61411764705882)
(0.42,4.86095238095238)
(0.52,4.33230769230769)
(0.22,3.78181818181818)
};
\addplot [purple, only marks, mark = otimes, thick] coordinates{
(0.888888888888889,2.05761316872428)
(0.222222222222222,0)
(0.9375,0.625)
(0.527777777777778,4.28849902534113)
(0.125,0)
(0.96,0.248888888888888)
(0.69,1.75304347826087)
(0.18,3.91111111111111)
(0.06,0)
(0.08,0)
};
\end{axis}
\end{tikzpicture}
\end{subfigure}

\vspace{0.5cm}
\begin{subfigure}{\textwidth}
\centering

\caption{Maximal distortion $\Lambda$}
\label{Fig3b}

\begin{tikzpicture}
\begin{axis}[
width = 1\textwidth, 
height = 0.6\textwidth,
xmajorgrids,
ymajorgrids,
xlabel = {Proportion of valid assignments},
xlabel style = {align=center, font=\small},
xmin = 0,
xmax = 1,
ylabel = {Distortion in \%},
ylabel style = {align=center, font=\small},
ymin = 0,
legend style = {font=\small,at={(0.02,-0.15)},anchor=north west,legend columns=3},
legend entries = {Order A-B-C$\qquad \qquad \quad \qquad \, \;$,Orders A-B-C and A-C-B$\qquad$,Order A-C-B,Orders A-C-B and B-C-A$\qquad$,Order B-C-A$\qquad \qquad \qquad \quad \;$,All orders$\quad \;$},
]
\addplot [blue, only marks, mark = square, thick] coordinates{
(0.555555555555556,13.3333333333333)
(0.75,2.77777777777778)
(0.625,5)
(0.4375,10.7142857142857)
(0.84,2.14285714285714)
(0.76,2.36842105263158)
(0.64,2.8125)
(0.37,6.48648648648649)
(0.68,3.52941176470588)
(0.52,4.61538461538462)
(0.22,10.9090909090909)
(0.36,8.88888888888889)
};
\addplot [red, only marks, mark = star, thick] coordinates{
(0.777777777777778,9.52380952380952)
(0.875,3.57142857142857)
(0.8125,5.76923076923077)
(0.333333333333333,12.5)
(0.92,2.60869565217391)
(0.88,2.72727272727273)
(0.55,5.45454545454545)
(0.84,3.80952380952381)
(0.42,8.57142857142857)
};
\addplot [black, only marks, mark = pentagon, thick] coordinates{
(0.625,15)
(0.76,6.31578947368421)
(0.68,9.41176470588235)
(0.52,18.4615384615385)
(0.22,16.3636363636364)
};
\addplot [orange, only marks, mark = diamond, thick] coordinates{
(0.555555555555556,26.6666666666667)
(0.75,8.33333333333333)
(0.4375,32.1428571428571)
(0.84,5.71428571428572)
(0.64,11.25)
(0.37,9.72972972972973)
(0.36,35.5555555555556)
};
\addplot [ForestGreen, only marks, mark = triangle, thick] coordinates{
(0.777777777777778,19.047619047619)
(0.875,10.7142857142857)
(0.8125,17.3076923076923)
(0.625,22.5)
(0.333333333333333,8.33333333333333)
(0.92,6.95652173913044)
(0.88,10.9090909090909)
(0.76,9.47368421052632)
(0.55,8.18181818181818)
(0.84,15.2380952380952)
(0.68,22.5882352941176)
(0.42,12.4285714285714)
(0.52,29.5384615384615)
(0.22,10.6363636363636)
};
\addplot [purple, only marks, mark = otimes, thick] coordinates{
(0.888888888888889,8.33333333333333)
(0.222222222222222,0)
(0.9375,5)
(0.527777777777778,6.57894736842105)
(0.125,0)
(0.96,3.33333333333334)
(0.69,4.78260869565217)
(0.18,9.16666666666667)
(0.06,0)
(0.08,0)
};
\end{axis}
\end{tikzpicture}

\end{subfigure}

\caption{Distortion of the Skip mechanism, theoretical model with $n \leq 5$}
\label{Fig3}

\end{figure}


The main message is reinforced by Figure~\ref{Fig3}, which shows the bias with the three draw orders as a function of how restrictive the constraint is.
Draw order A-B-C never has a greater average bias than draw order A-C-B, and draw order A-C-B never has a greater average bias than draw order B-C-A. In other words, the dots are vertically aligned, but two or three draw orders have sometimes the same distortion, as well as some sets of parameters lead to close or equal proportion of valid assignments.
This pattern almost holds with respect to maximal bias, too, except for the two sets of parameters discussed above.

Maximal distortion $\Lambda$ may be close to or even exceed 30\% if one of the unfavourable draw orders A-C-B and B-C-A is chosen, while it always remains below 13.5\% with the draw order A-B-C.
On the other hand, no clear relationship exists between the proportion of valid assignments and the size of distortions. Nonetheless, the most worrying cases seem to occur if about half of all assignments violate the constraint.

To summarise, a draw order where the first two pots contain more teams affected by the constraint is preferred by the natural measure of average distortion $\Delta$. This ranking of draw orders also emerges with respect to maximal distortion $\Lambda$ in almost all cases, at least if there are $n \leq 5$ teams in each of the three pots.

\section{Field evidence: real-world case studies} \label{Sec5}

This section considers three draws of sports tournaments, where the organiser
(1) imposed a constraint that is analogous to the one in the theoretical model of Section~\ref{Sec4}; and
(2) used the Skip mechanism with the ``natural'' draw order from the strongest to the weakest pot.
In particular, Section~\ref{Sec51} examines the 2018 FIFA World Cup draw, Section~\ref{Sec52} studies the draw of the European Qualifiers for the 2022 FIFA World Cup, while Section~\ref{Sec53} analyses the 2019 FIBA Basketball World Cup draw, which consisted of two separate draws in our setting.
In this section, the pots are denoted by numbers as in the associated tournaments.

In Sections~\ref{Sec51} and \ref{Sec52}, the assignment probabilities cannot be calculated exactly, neither for a uniform draw, nor for the Skip mechanism, because the set of valid assignments is prohibitively large.
Uniform distribution can be approximated by a rejection sampler \citep[Section~2.1]{RobertsRosenthal2024}: an unconstrained assignment is generated without taking any draw restriction into account, which is retained if all constraints are satisfied and dismissed if any constraint is violated.
The Skip mechanism is implemented by the backtracking algorithm given in \citet{Csato2025c} and \citet{RobertsRosenthal2024}.

However, considering all possible orders of the teams is impossible. In the 2018 FIFA World Cup draw, there exist $\left( 8! \right)^3 \cdot 7! \approx 3.3 \cdot 10^{17}$ different orders in which the teams can be drawn from the four pots. The number of permutations, $\left( 10! \right)^5 \cdot 5! \approx 7.55 \cdot 10^{34}$, is even much higher for the second case study. Therefore, the distortions are calculated on the basis of 1 million (Section~\ref{Sec51}) and 5 million (Section~\ref{Sec52}) simulation runs, respectively.

Even this is quite expensive computationally, the running times reach several hours on a standard high-performance laptop for each draw order. Therefore, although all ($4!= 24$) draw orders are considered in Section~\ref{Sec51}, Section~\ref{Sec52} reports the results only for 11 draw orders out of the $6! = 720$ possible draw orders implied by the six pots.

On the other hand, the example of Section~\ref{Sec53} involves two draws with four pots containing four teams each. Similar to Section~\ref{Sec42}, this allows to investigate all possible cases and calculate the exact assignment probabilities: the rejection sampler should check $\left( 4! \right)^3 = 13{,}824$ scenarios, and the number of orders in which the teams can be drawn is $\left( 4! \right)^4 = 331{,}776$ for each of the $4! = 24$ versions of the Skip mechanism. This requires only a few minutes to determine the probabilities for any draw order of the pots.

\subsection{2018 FIFA World Cup draw} \label{Sec51}

The 2018 FIFA World Cup was contested by 32 teams, divided into four pots of eight teams each. For the first time, the pots were determined by the strengths of the teams, measured by the FIFA World Ranking of October 2017. The host Russia was automatically assigned to the strongest Pot 1.

\begin{table}[t!]
  \centering
  \caption{Seeding in the 2018 FIFA World Cup draw}
  \label{Table3}
     \rowcolors{1}{}{gray!20}
\centerline{
\begin{threeparttable}
    \begin{tabularx}{0.8\textwidth}{l CCCC c} \toprule \hiderowcolors
    Confederation      & Pot 1 & Pot 2 & Pot 3 & Pot 4 & Prohibited pairs \\ \bottomrule \showrowcolors
    AFC   & ---   & ---   & 1     & 4     & 4 \\
    CAF   & ---   & ---   & 3     & 2     & 6 \\
    CONCACAF & ---   & 1     & 1     & 1     & 3 \\
    CONMEBOL & 2     & 3     & ---   & ---   & 6 \\
    UEFA  & 5+1   & 4     & 3     & 1     & --- \\ \toprule
    \end{tabularx}
\begin{tablenotes} \footnotesize
\item
\emph{Notes:} The cells indicate the number of teams in the given pot from the different confederations. \\
+1 refers to the host Russia since its automatic assignment to Group A influences assignment probabilities. \\
The column Prohibited pairs shows the number of team pairs that cannot be assigned to the same group.
\end{tablenotes}
\end{threeparttable} 
}   
\end{table}

The draw constraints are as follows \citep{FIFA2017c}:
\begin{itemize}
\item
Russia is assigned to Group A;
\item
No group could have more than one team from the same confederation except for UEFA (AFC, CAF, CONCACAF, CONMEBOL);
\item
Each group should have at least one but no more than two European (UEFA) teams. 
\end{itemize}
The implied restrictions are summarised in Table~\ref{Table3}; for instance, two South American teams in Pot 1 and three in Pot 3 mean that six pairs of national teams cannot be in the same group. Hence, there are $6 \cdot 8 \cdot 8 - 19 = 365$ nonzero assignment probabilities.

The draw used the Skip mechanism to ensure that all draw constraints were satisfied \citep{FIFA2017c, Guyon2018d}.
The draw sequence started with Pot 1 and ended with Pot 4.

\begin{table}[t!]
  \centering
  \caption{The effect of draw order on the Skip mechanism, 2018 FIFA World Cup draw}
  \label{Table4}
     \rowcolors{1}{}{gray!20}
\centerline{
\begin{threeparttable}
    \begin{tabularx}{0.8\textwidth}{l CCCC c} \toprule \hiderowcolors
    Draw order & Average bias & Maximal bias & Maximal 10 bias \\ \bottomrule \showrowcolors
    1-2-3-4 & 0.824\% & 10.29\% & 6.24\% \\
    1-2-4-3 & 0.855\% & \textcolor{white}{0}9.97\% & 5.30\% \\
    1-3-2-4 & 1.060\% & \textcolor{gray!20}{0}8.46\% & 5.64\% \\
    1-3-4-2 & 1.037\% & \textcolor{white}{0}5.19\% & 4.40\% \\
    1-4-2-3 & 0.922\% & \textcolor{gray!20}{0}7.35\% & 5.02\% \\
    1-4-3-2 & 1.029\% & \textcolor{white}{0}6.33\% & 4.66\% \\ \hline
    2-1-3-4 & 0.850\% & 10.60\% & 6.36\% \\
    2-1-4-3 & 0.878\% & \textcolor{white}{0}9.96\% & 5.28\% \\
    2-3-1-4 & 1.417\% & \textcolor{gray!20}{0}6.08\% & 5.71\% \\
    2-3-4-1 & 1.322\% & \textcolor{white}{0}6.02\% & 5.46\% \\
    2-4-1-3 & 1.044\% & \textcolor{gray!20}{0}7.45\% & 5.56\% \\
    2-4-3-1 & 1.155\% & \textcolor{white}{0}7.53\% & 5.20\% \\ \hline
    3-1-2-4 & 1.053\% & \textcolor{gray!20}{0}8.47\% & 5.62\% \\
    3-1-4-2 & 1.034\% & \textcolor{white}{0}5.17\% & 4.41\% \\
    3-2-1-4 & 1.268\% & \textcolor{gray!20}{0}5.71\% & 5.33\% \\
    3-2-4-1 & 1.205\% & \textcolor{white}{0}5.48\% & 5.17\% \\
    3-4-1-2 & 1.275\% & 16.38\% & 6.50\% \\
    3-4-2-1 & 1.539\% & 12.77\% & 7.42\% \\ \hline
    4-1-2-3 & 0.915\% & \textcolor{gray!20}{0}7.33\% & 4.97\% \\
    4-1-3-2 & 1.057\% & \textcolor{white}{0}6.38\% & 4.67\% \\
    4-2-1-3 & 0.960\% & \textcolor{gray!20}{0}6.94\% & 4.99\% \\
    4-2-3-1 & 1.134\% & \textcolor{white}{0}6.76\% & 4.67\% \\
    4-3-1-2 & 1.276\% & 15.27\% & 6.06\% \\
    4-3-2-1 & 1.554\% & 11.76\% & 6.97\% \\ \bottomrule
    \end{tabularx}
\begin{tablenotes} \footnotesize
\item
\emph{Notes:} The column Maximal 10 bias shows the mean of the 10 highest distortions. \\
Draw order 1-2-3-4 is both the official and the recommended.
\end{tablenotes}
\end{threeparttable} 
}   
\end{table}

\citet{Csato2025c} computed the distortion of the Skip mechanism with the $4!= 24$ possible draw orders of the four pots as can be seen in Table~\ref{Table4}.
The results are fully in line with the theoretical model of Section~\ref{Sec4}, despite the existence of one additional pot and 19 excluded pairs.
First, exchanging the first two pots in the draw order has a moderated effect on the bias from a uniformly distributed draw since the constraint for the European teams does not depend on it.
Second, the average distortion is smaller if pots with more European teams are drawn first. For example, the four smallest values of $\Delta$ occur if the draw starts with Pots 1 and 2. While this choice is not optimal with respect to the maximal and maximal 10 distortions, the highest values usually appear if the draw starts with Pots 3 and 4. The average bias $\Delta$ is also maximal if the draw ends with Pots 2 and 1 in this order.
Third, the distortions for two draw orders that differ only in the order of the first two pots are strongly correlated.  
Last but not least, the order of the pots has a powerful effect on the bias of the draw: the optimal draw order reduces average distortion $\Delta$ by 47\% compared to the worst version of the Skip mechanism, although the proportion of UEFA teams is only $14/32 = 43.75$\%.

To summarise, the distortions of the 2018 FIFA World Cup draw could not have been reduced by a straightforward reform in the Skip mechanism. While \citet{Csato2025c} has already reached this conclusion based on average bias values, now we are able to explain the outcome of these black box calculations.

\subsection{European Qualifiers for the 2022 FIFA World Cup} \label{Sec52}

In the first round of the European section of the 2022 FIFA World Cup qualification, 55 national teams were drawn into five groups of five teams and five groups of six teams each. The teams were classified into six pots based on the FIFA World Rankings of November 2020.
The group winners qualified for the 2022 FIFA World Cup, while the runners-up progressed to the second round of the qualification.

\begin{table}[t!]
  \centering
  \caption{Draw constraints in the European Qualifiers for the 2022 FIFA World Cup}
  \label{Table5}
  \rowcolors{1}{}{gray!20}
\begin{threeparttable}
    \begin{tabularx}{0.8\textwidth}{l CCC CCC} \toprule
    Type of restriction & Pot 1 & Pot 2 & Pot 3 & Pot 4 & Pot 5 & Pot 6 \\  \bottomrule
    Prohibited clash & ---   & 2     & 3     & 1     & 4     & --- \\
    Winter venue & ---   & 1     & 4     & 1     & 4     & --- \\
    Excessive travel & 5     & 1     & 9     & 2     & 16    & 3 \\ \toprule
    \end{tabularx}
\begin{tablenotes} \footnotesize
\item
\emph{Notes:} Prohibited clash and excessive travel are taken into account for both pots. \\
The pair Faroe Islands and Iceland is counted as a prohibited clash.
\end{tablenotes}
\end{threeparttable}
\end{table}

The draw constraints are as follows \citep{UEFA2020c}:
\begin{itemize}
\item
\emph{Competition-related reasons}:
The four participants in the UEFA Nations League Finals 2021 should play in a group of five teams (Groups A--E).
\item
\emph{Prohibited team clashes}:
Four pairs of teams cannot be drawn into the same group due to political reasons.
\item
\emph{Winter venue restrictions}:
At most two out of the ten countries identified as venues with a risk of severe winter conditions can be placed in a group. The Faroe Islands and Iceland, having the highest risk, cannot play in the same group.
\item
\emph{Excessive travel restrictions}:
A maximum of one country pair out of the 18 exhibiting excessive travel relations can be drawn into the same group.
\end{itemize}
Table~\ref{Table5} reports the number of draw constraints in each pot.

The draw used the Skip mechanism, starting with Pot 1 and continuing with Pot 2 until Pot 6. Each pot was emptied before proceeding to the next pot. The five teams from Pot 6 were assigned to the last five groups (Groups F--J).

Based on Section~\ref{Sec4}, the draw is worth starting with the pots containing more countries that are affected by the winter venue constraint; however, this consideration does not apply to pots without such a team (Pot 1, Pot 6). In addition, Pot 5 contains Kazakhstan, which is by far the most important country with respect to excessive travel restrictions: it appears in 13 pairs, while the second is Iceland (Pot 3) with only six restrictions. If Kazakhstan is drawn at the beginning, a distant country can be placed into its group relatively early, which excludes the assignment of another team with such a property. If Kazakhstan is drawn at the end, two distant countries are more likely to be in the same group, which does not allow the assignment of Kazakhstan into this group. Thus, the best strategy seems to be keeping Kazakhstan, that is, Pot 5, in the middle of the draw order.

\begin{table}[t!]
  \centering
  \caption{The effect of draw order on the Skip mechanism, \\ European Qualifiers for the 2022 FIFA World Cup draw}
  \label{Table6}
     \rowcolors{1}{gray!20}{}
\centerline{
\begin{threeparttable}
    \begin{tabularx}{0.8\textwidth}{l CCCC c} \toprule \hiderowcolors
    Draw order & Average bias & Maximal bias & Maximal 10 bias \\ \bottomrule \showrowcolors
    1-2-3-4-5-6 & 0.202\% & 3.80\% & 3.02\% \\
    1-3-5-2-4-6 & 0.141\% & 2.69\% & 1.99\% \\
    1-3-5-6-4-2 & 0.142\% & 2.39\% & 1.89\% \\
    1-3-6-5-2-4 & 0.133\% & 2.14\% & 1.59\% \\
    1-6-3-5-2-4 & 0.134\% & 2.15\% & 1.59\% \\
    1-5-3-2-4-6 & 0.198\% & 4.82\% & 3.12\% \\
    1-5-3-6-4-2 & 0.199\% & 4.82\% & 3.04\% \\
    1-2-4-6-3-5 & 0.194\% & 4.20\% & 2.72\% \\
    1-6-4-2-3-5 & 0.194\% & 4.18\% & 2.72\% \\
    1-2-4-6-5-3 & 0.188\% & 4.18\% & 2.56\% \\
    1-6-4-2-5-3 & 0.188\% & 4.17\% & 2.56\% \\ \toprule
    \end{tabularx}
\begin{tablenotes} \footnotesize
\item
\emph{Notes:} The column Maximal 10 bias shows the mean of the 10 highest distortions. \\
Draw order 1-2-3-4-5-6 is the official. \\
Draw order 1-3-6-5-2-4 is the recommended.
\end{tablenotes}
\end{threeparttable} 
}   
\end{table}

Table~\ref{Table6} uncovers the distortion of the Skip mechanism with the original (1-2-3-4-5-6) and ten other draw orders. Now exchanging Pots 3 and 5 strongly affects aggregated distortions due to excessive travel relations, while the order of Pots 2, 4, 6 only moderately influences them. The recommendation of retaining Pot 5 in a central position is important. The official draw procedure is far from being optimal, average distortion $\Lambda$ can be decreased by 34\% and the largest distortions by almost 50\% if the draw order 1-3-6-5-2-4 is chosen. Nonetheless, the average bias $\Delta$ is only about a fifth of the 2018 FIFA World Cup draw, and the maximal bias $\Lambda$ is more than halved due to a smaller proportion of restrictions.

\begin{figure}[t!]
\centering

\begin{tikzpicture}
\begin{axis}[
name = axis1,
width = 0.5\textwidth, 
height = 0.8\textwidth,
title = {Average distortion $\Delta$},
title style = {align=center, font=\small},
xmajorgrids = true,
ymajorgrids = true,
xmin = 0,
xmax = 1.15,
scaled x ticks = false,
xlabel = {Average distortion (\%)},
xlabel style = {align=center, font=\small},
xticklabel style = {/pgf/number format/fixed,/pgf/number format/precision=5},
ytick style = {draw = none},
ymin = 0,
ymax = 56,
y dir = reverse,
ylabel = {Rank of the national team in the draw},
ylabel style = {align=center, font=\small},
legend style = {font=\small,at={(0,-0.1)},anchor=north west,legend columns=4},
legend entries = {Draw order 1-2-3-4-5-6$\qquad$, Draw order 1-3-6-5-2-4$\qquad$, Draw order 1-5-3-2-4-6}
]
\addplot [red, only marks, mark = star, thick] coordinates{
(0.154977,1)
(0.1605515,2)
(0.101859111111111,3)
(0.198960222222222,4)
(0.162848,5)
(0.156964,6)
(0.0591066666666666,7)
(0.0676222222222223,8)
(0.060504,9)
(0.0625677777777778,10)
(0.0950026666666667,11)
(0.112846222222222,12)
(0.0913195555555555,13)
(0.0938011111111111,14)
(0.0977566666666665,15)
(0.759301363636364,16)
(0.160493181818182,17)
(0.0980064444444445,18)
(0.0949155555555555,19)
(0.0938131111111111,20)
(0.332559534883721,21)
(0.246064888888889,22)
(0.260311333333333,23)
(0.246878,24)
(0.366835333333333,25)
(0.256719333333333,26)
(0.660774090909091,27)
(0.260721555555556,28)
(0.249460666666667,29)
(0.365244,30)
(0.170466818181818,31)
(0.101089777777778,32)
(0.102941333333333,33)
(0.103429777777778,34)
(0.103318888888889,35)
(0.103168222222222,36)
(0.261507333333333,37)
(0.666958666666666,38)
(0.259925111111111,39)
(0.103790222222222,40)
(0.226781333333333,41)
(0.226107555555556,42)
(0.225066136363636,43)
(0.180992888888889,44)
(0.172886222222222,45)
(0.177627619047619,46)
(0.545456888888889,47)
(0.171193111111111,48)
(0.168188888888889,49)
(0.336371555555556,50)
(0.0793152173913044,51)
(0.142947826086957,52)
(0.140705217391305,53)
(0.092873043478261,54)
(0.142936956521739,55)
};
\addplot [blue, only marks, mark = square, thick] coordinates{
(0.18432,1)
(0.1632435,2)
(0.0773704444444445,3)
(0.131108222222222,4)
(0.1669235,5)
(0.182707,6)
(0.105204222222222,7)
(0.104652,8)
(0.102734444444444,9)
(0.105676444444444,10)
(0.0338342222222222,11)
(0.0390453333333333,12)
(0.0336637777777778,13)
(0.0315102222222222,14)
(0.0324926666666667,15)
(0.200315909090909,16)
(0.0864399999999999,17)
(0.0332602222222222,18)
(0.0314275555555556,19)
(0.0343939999999999,20)
(0.106030465116279,21)
(0.136686444444444,22)
(0.145642,23)
(0.140408,24)
(0.247956,25)
(0.144599777777778,26)
(0.347119545454545,27)
(0.146047555555556,28)
(0.141206666666667,29)
(0.244105333333333,30)
(0.0916254545454545,31)
(0.0429595555555555,32)
(0.0395944444444445,33)
(0.0404455555555555,34)
(0.041570888888889,35)
(0.0407146666666666,36)
(0.0484293333333332,37)
(0.250449333333333,38)
(0.0499248888888887,39)
(0.0418544444444444,40)
(0.161523333333333,41)
(0.161722888888889,42)
(0.182098863636363,43)
(0.168172222222222,44)
(0.196774,45)
(0.201420238095238,46)
(0.475177111111111,47)
(0.199024666666667,48)
(0.198414888888889,49)
(0.197180666666667,50)
(0.127786521739131,51)
(0.181787826086957,52)
(0.180813043478261,53)
(0.174731304347826,54)
(0.178846521739131,55)
};
\addplot [ForestGreen, only marks, mark = triangle, very thick] coordinates{
(0.2528925,1)
(0.2220195,2)
(0.401922222222222,3)
(0.509814222222222,4)
(0.2171765,5)
(0.25329,6)
(0.203844666666667,7)
(0.206441777777778,8)
(0.205022666666667,9)
(0.201740666666667,10)
(0.0406746666666665,11)
(0.188260222222222,12)
(0.0369591111111111,13)
(0.039808,14)
(0.0343431111111112,15)
(0.192290681818182,16)
(0.110474772727273,17)
(0.0397984444444445,18)
(0.0426133333333333,19)
(0.0382097777777779,20)
(0.220749534883721,21)
(0.174987111111111,22)
(0.197782888888889,23)
(0.177863333333333,24)
(0.285818,25)
(0.196763111111111,26)
(0.440147272727273,27)
(0.196835111111111,28)
(0.173942666666667,29)
(0.285715555555556,30)
(0.106702727272727,31)
(0.0434595555555555,32)
(0.0386766666666667,33)
(0.0364726666666667,34)
(0.0405675555555556,35)
(0.0403053333333334,36)
(0.0605055555555554,37)
(0.243632666666666,38)
(0.0608339999999999,39)
(0.0416513333333333,40)
(0.202773333333333,41)
(0.203968888888889,42)
(0.256825909090909,43)
(0.234569111111111,44)
(0.241233111111111,45)
(0.276769523809524,46)
(0.992450888888889,47)
(0.242826666666667,48)
(0.238408444444444,49)
(0.251392222222222,50)
(0.246294347826087,51)
(0.252227391304348,52)
(0.251007391304348,53)
(0.276325217391304,54)
(0.251060869565217,55)
};

\node at (1,5.25) {Pot 1};
\draw [black,thick,dashed] (\pgfkeysvalueof{/pgfplots/xmin},10.5) -- (\pgfkeysvalueof{/pgfplots/xmax},10.5);
\node at (1,15.5) {Pot 2};
\draw [black,thick,dashed] (\pgfkeysvalueof{/pgfplots/xmin},20.5) -- (\pgfkeysvalueof{/pgfplots/xmax},20.5);
\node at (1,25.5) {Pot 3};
\draw [black,thick,dashed] (\pgfkeysvalueof{/pgfplots/xmin},30.5) -- (\pgfkeysvalueof{/pgfplots/xmax},30.5);
\node at (1,35.5) {Pot 4};
\draw [black,thick,dashed] (\pgfkeysvalueof{/pgfplots/xmin},40.5) -- (\pgfkeysvalueof{/pgfplots/xmax},40.5);
\node at (1,45.5) {Pot 5};
\draw [black,thick,dashed] (\pgfkeysvalueof{/pgfplots/xmin},50.5) -- (\pgfkeysvalueof{/pgfplots/xmax},50.5);
\node at (1,52.75) {Pot 6};
\end{axis}

\begin{axis}[
at = {(axis1.south east)},
xshift = 0.1\textwidth,
width = 0.5\textwidth, 
height = 0.8\textwidth,
title = {Maximal distortion $\Lambda$},
title style = {align=center, font=\small},
xmajorgrids = true,
ymajorgrids = true,
xmin = 0,
xlabel = {Maximal distortion (\%)},
scaled x ticks = false,
xlabel style = {align=center, font=\small},
xticklabel style = {/pgf/number format/fixed,/pgf/number format/precision=4},
ytick style = {draw = none},
ymin = 0,
ymax = 56,
ylabel = {Rank of the national team in the draw},
ylabel style = {align=center, font=\small},
y dir = reverse,
]
\addplot [red, only marks, mark = star, thick] coordinates{
(1.11572,1)
(1.07822,2)
(1.45664,3)
(1.30993,4)
(1.0931,5)
(1.09859,6)
(0.40328,7)
(0.41653,8)
(0.394220000000001,9)
(0.405029999999999,10)
(0.49003,11)
(0.488420000000001,12)
(0.431720000000001,13)
(0.450149999999999,14)
(0.45877,15)
(3.50982,16)
(1.17567,17)
(0.467970000000001,18)
(0.44483,19)
(0.459670000000001,20)
(3.80441,21)
(1.74644,22)
(1.71891,23)
(1.77891,24)
(3.50982,25)
(1.72927,26)
(3.50083,27)
(1.73617,28)
(1.79054,29)
(3.48959,30)
(1.4147,31)
(1.15265,32)
(1.16449,33)
(1.19266,34)
(1.13642,35)
(1.18178,36)
(3.32658,37)
(3.80441,38)
(3.3041,39)
(1.16127,40)
(0.984249999999999,41)
(0.990749999999999,42)
(1.21008,43)
(1.19161,44)
(1.11036,45)
(0.92805,46)
(2.40841,47)
(1.08886,48)
(1.09953,49)
(1.36983,50)
(0.403100000000001,51)
(2.40841,52)
(2.36606,53)
(0.52276,54)
(2.38919,55)
};
\addplot [blue, only marks, mark = square, thick] coordinates{
(1.2635,1)
(0.94239,2)
(0.32667,3)
(1.45496,4)
(0.95533,5)
(1.23281,6)
(0.986400000000001,7)
(0.99384,8)
(0.976720000000002,9)
(0.984079999999998,10)
(0.10346,11)
(0.112620000000001,12)
(0.10366,13)
(0.0966999999999996,14)
(0.110880000000001,15)
(0.74816,16)
(0.865579999999999,17)
(0.0993399999999992,18)
(0.0924800000000003,19)
(0.12041,20)
(0.58647,21)
(0.53315,22)
(0.481989999999999,23)
(0.529539999999999,24)
(1.18383,25)
(0.47543,26)
(2.1354,27)
(0.48311,28)
(0.51132,29)
(1.19509,30)
(0.865579999999999,31)
(0.12282,32)
(0.12528,33)
(0.113229999999999,34)
(0.12848,35)
(0.118170000000001,36)
(0.30931,37)
(0.74816,38)
(0.326439999999999,39)
(0.118370000000001,40)
(0.952550000000001,41)
(0.92723,42)
(1.19509,43)
(0.746520000000001,44)
(0.98413,45)
(2.1354,46)
(1.92341,47)
(0.97136,48)
(0.977690000000001,49)
(1.72314,50)
(0.51434,51)
(1.92341,52)
(1.89682,53)
(1.45496,54)
(1.88578,55)
};
\addplot [ForestGreen, only marks, mark = triangle, very thick] coordinates{
(3.64806,1)
(2.24247,2)
(4.81735,3)
(4.67068,4)
(2.20808,5)
(3.61272,6)
(1.65683,7)
(1.66548,8)
(1.68735,9)
(1.66814,10)
(0.149619999999999,11)
(0.60757,12)
(0.13015,13)
(0.156009999999999,14)
(0.13025,15)
(0.71963,16)
(0.86502,17)
(0.156669999999999,18)
(0.15888,19)
(0.146109999999999,20)
(0.79356,21)
(0.523410000000001,22)
(0.625089999999999,23)
(0.51833,24)
(1.16281,25)
(0.628859999999999,26)
(2.29102,27)
(0.614830000000001,28)
(0.50239,29)
(1.1752,30)
(0.86502,31)
(0.12973,32)
(0.127279999999999,33)
(0.12375,34)
(0.15372,35)
(0.116579999999999,36)
(0.33474,37)
(0.71963,38)
(0.383009999999999,39)
(0.137469999999999,40)
(0.83993,41)
(0.801349999999999,42)
(1.1752,43)
(1.9247,44)
(0.98554,45)
(2.29102,46)
(4.81735,47)
(0.990390000000001,48)
(0.993230000000001,49)
(1.85736,50)
(1.59476,51)
(2.57554,52)
(2.58753,53)
(1.64665,54)
(2.55287,55)
};

\node at (4.5,5.25) {Pot 1};
\draw [black,thick,dashed] (\pgfkeysvalueof{/pgfplots/xmin},10.5) -- (\pgfkeysvalueof{/pgfplots/xmax},10.5);
\node at (4.5,15.5) {Pot 2};
\draw [black,thick,dashed] (\pgfkeysvalueof{/pgfplots/xmin},20.5) -- (\pgfkeysvalueof{/pgfplots/xmax},20.5);
\node at (4.5,25.5) {Pot 3};
\draw [black,thick,dashed] (\pgfkeysvalueof{/pgfplots/xmin},30.5) -- (\pgfkeysvalueof{/pgfplots/xmax},30.5);
\node at (4.5,35.5) {Pot 4};
\draw [black,thick,dashed] (\pgfkeysvalueof{/pgfplots/xmin},40.5) -- (\pgfkeysvalueof{/pgfplots/xmax},40.5);
\node at (4.5,45.5) {Pot 5};
\draw [black,thick,dashed] (\pgfkeysvalueof{/pgfplots/xmin},50.5) -- (\pgfkeysvalueof{/pgfplots/xmax},50.5);
\node at (4.5,52.75) {Pot 6};
\end{axis}
\end{tikzpicture}

\caption{The bias of the Skip mechanism with different draw orders for the \\ national teams in the European Qualifiers for the 2022 FIFA World Cup}
\label{Fig4}

\end{figure}


Finally, Figure~\ref{Fig4} considers the distortion of the draw at the level of national teams for three variants of the Skip mechanism. Draw order 1-3-6-5-2-4 implies smaller distortions for Pots 2, 3, 4, which is beneficial since these teams are stronger and have a higher probability of qualifying than those in Pots 5 and 6. On the other hand, draw order 1-5-3-2-4-6 is a bad option---even though it is better than the original draw order according to average distortion---as it increases the biases for the best 10 teams in Pot 1.

To conclude, a careful choice of the order of the pots based on theoretical ideas can reduce the distortions of the assignment probabilities even if checking all possible draw orders is extremely costly---note that the six pots have $6! = 720$ permutations. Furthermore, different draw orders may be optimal for different subsets of the teams.

\subsection{2019 FIBA Basketball World Cup draw} \label{Sec53}

The 2019 FIBA Basketball World Cup was the first edition of this competition with 32 teams. The host China and the three strongest teams based on the 2019 FIBA World Ranking were allocated to Pot 1. Pots 2--8 contained the remaining 28 teams according to their strength. The first set of groups--- Groups A, C, E and G---contained one team from Pots 1, 4, 5 and 8, respectively, while the second set---Groups B, D, F and H---contained one team from Pots 2, 3, 6 and 7, respectively.
Even though China was automatically assigned to Group A and the United States to Group E, we ignore these constraints that may cause additional distortions \citep{Csato2026a}, since these constraints can be easily ensured by labelling the groups only \emph{after} the draw.

Analogous to the 2018 FIFA World Cup, two teams from the same qualification zone could not be drawn into the same group except for Europe. The number of European teams is allowed to be between one and two in each group. Canada was moved from Pot 5 to Pot 6 and replaced with the best team of Pot 6, Iran, in order to avoid two teams from the Americas playing in the same group.

\begin{table}[t!]
  \centering
  \caption{Seeding in the 2019 FIBA World Cup draw}
  \label{Table7}

\begin{subtable}{\textwidth}
  \centering
  \caption{Groups A, C, E, and G}
  \label{Table7a}
    \rowcolors{1}{}{gray!20}
    \begin{tabularx}{0.8\textwidth}{l CCCC c} \toprule \hiderowcolors
    Qualification zone & Pot 1 & Pot 4 & Pot 5 & Pot 8 & Prohibited pairs \\ \bottomrule \showrowcolors
    Africa & ---   & ---   & ---   & 2     & --- \\
    Americas & 1     & 3     & ---   & ---   & 3 \\
    Asia  & 1     & ---   & 1     & 2     & 5 \\
    Europe & 2     & 1     & 3     & ---   & --- \\ \toprule
    \end{tabularx}
\end{subtable}

\vspace{0.5cm}
\begin{subtable}{\textwidth}
  \centering
  \caption{Groups B, D, F, and H}
  \label{Table7b}  
\begin{threeparttable}
    \rowcolors{1}{}{gray!20}
    \begin{tabularx}{0.8\textwidth}{l CCCC c} \toprule \hiderowcolors
    Qualification zone & Pot 2 & Pot 3 & Pot 6 & Pot 7 & Prohibited pairs \\ \bottomrule \showrowcolors
    Africa & ---   & ---   & ---   & 3     & --- \\
    Americas & 1     & 1     & 1     & ---   & 3 \\
    Asia  & ---   & 1     & 2     & 1     & 5 \\
    Europe & 3     & 2     & 1     & ---   & --- \\ \toprule
    \end{tabularx}
\begin{tablenotes} \footnotesize
\item
\emph{Notes:} The cells indicate the number of teams in the given pot from the different qualification zones. \\
The column Prohibited pairs shows the number of team pairs that cannot be assigned to the same group.
\end{tablenotes}
\end{threeparttable} 
\end{subtable}
 
\end{table}

The implied restrictions are summarised in Table~\ref{Table7}. Indeed, there are $6 \cdot 4 \cdot 4 - 8 = 88$ nonzero assignment probabilities for the second set of groups. But this number is only $85$ for the first set of groups because Turkey, the only non-American team in Pot 4, should play against the United States from Pot 1---thus, it cannot be assigned to a group with the three other countries in Pot 1 (China, France, Spain).
The draw used the Skip mechanism, and the draw sequence started from Pots 1 and 2 in increasing order \citep{FIBA2019}.

\begin{table}[t!]
  \centering
  \caption{The effect of draw order on the Skip mechanism in the \\ 2019 FIBA Basketball World Cup draw I.~(Groups A, C, E, G)}
  \label{Table8}
     \rowcolors{1}{}{gray!20}
\centerline{
\begin{threeparttable}
    \begin{tabularx}{0.8\textwidth}{l CCCC c} \toprule \hiderowcolors
    Draw order & Average bias & Maximal bias & Maximal 8 bias \\ \bottomrule \showrowcolors
    1-4-5-8 & 0\%   & 0\%   & 0\% \\
    1-4-8-5 & 0\%   & 0\%   & 0\% \\
    1-5-4-8 & 0\%   & 0\%   & 0\% \\
    1-5-8-4 & 0\%   & 0\%   & 0\% \\
    1-8-4-5 & 0\%   & 0\%   & 0\% \\
    1-8-5-4 & 0\%   & 0\%   & 0\% \\ \hline
    4-1-5-8 & 0\%   & 0\%   & 0\% \\
    4-1-8-5 & 0\%   & 0\%   & 0\% \\
    4-5-1-8 & 1.569\% & \textcolor{gray!20}{0}8.33\% & \textcolor{gray!20}{0}5.21\% \\
    4-5-8-1 & 1.569\% & \textcolor{white}{0}8.33\% & \textcolor{white}{0}5.21\% \\
    4-8-1-5 & 3.137\% & 16.67\% & 10.42\% \\
    4-8-5-1 & 3.137\% & 16.67\% & 10.42\% \\ \hline
    5-1-4-8 & 0\%   & 0\%   & 0\% \\
    5-1-8-4 & 0\%   & 0\%   & 0\% \\
    5-4-1-8 & 1.569\% & \textcolor{gray!20}{0}8.33\% & \textcolor{gray!20}{0}5.21\% \\
    5-4-8-1 & 1.569\% & \textcolor{white}{0}8.33\% & \textcolor{white}{0}5.21\% \\
    5-8-1-4 & 0\%   & 0\%   & 0\% \\
    5-8-4-1 & 0.196\% & \textcolor{white}{0}1.04\% & \textcolor{white}{0}0.65\% \\ \hline
    8-1-4-5 & 0\%   & 0\%   & 0\% \\
    8-1-5-4 & 0\%   & 0\%   & 0\% \\
    8-4-1-5 & 3.137\% & 16.67\% & 10.42\% \\
    8-4-5-1 & 3.137\% & 16.67\% & 10.42\% \\
    8-5-1-4 & 0\%   & 0\%   & 0\% \\
    8-5-4-1 & 0.196\% & \textcolor{white}{0}1.04\% & \textcolor{white}{0}0.65\% \\ \bottomrule
    \end{tabularx}
\begin{tablenotes} \footnotesize
\item
\emph{Notes:} The column Maximal 8 bias shows the mean of the eight highest distortions. \\
Draw order 1-4-5-8 is both the official and the recommended.
\end{tablenotes}
\end{threeparttable} 
}   
\end{table}

Table~\ref{Table8} presents the distortions of the Skip mechanism for the first set of groups under the $4! = 24$ possible draw orders. The draw is uniform under 14 draw orders, including the official method and all starting with Pot 1. Two draw orders imply the same biases if the first two pots coincide in them. In the four most biased draw procedures, Pot 4 with one European team is drawn before Pots 1 and 5 that contain more European teams. While the worst maximal bias is slightly higher than for the 2018 FIFA World Cup in Table~\ref{Table4}, the worst average bias is more than doubled and is above three percentage points.
On the other hand, if the draw sequence contains Pot 5, Pot 1, and Pot 4 in this order, then the Skip mechanism is equivalent to a uniform draw. Consequently, the theoretical results of Section~\ref{Sec4} remain valid in this example, too. 

\begin{table}[t!]
  \centering
  \caption{The effect of draw order on the Skip mechanism in the \\ 2019 FIBA Basketball World Cup draw II.~(Groups B, D, F, H)}
  \label{Table9}
     \rowcolors{1}{}{gray!20}
\centerline{
\begin{threeparttable}
    \begin{tabularx}{0.8\textwidth}{l CCCC c} \toprule \hiderowcolors
    Draw order & Average bias & Maximal bias & Maximal 8 bias \\ \bottomrule \showrowcolors
    2-3-6-7 & 1.815\% & \textcolor{gray!20}{0}8.51\% & \textcolor{gray!20}{0}6.90\% \\
    2-3-7-6 & 1.539\% & \textcolor{white}{0}8.33\% & \textcolor{white}{0}5.86\% \\
    2-6-3-7 & 2.588\% & 16.67\% & \textcolor{gray!20}{0}9.51\% \\
    2-6-7-3 & 2.320\% & 16.67\% & \textcolor{white}{0}8.51\% \\
    2-7-3-6 & 0.260\% & \textcolor{gray!20}{0}1.04\% & \textcolor{gray!20}{0}1.04\% \\
    2-7-6-3 & 2.095\% & \textcolor{white}{0}9.38\% & \textcolor{white}{0}8.20\% \\ \hline
    3-2-6-7 & 1.827\% & \textcolor{gray!20}{0}8.51\% & \textcolor{gray!20}{0}6.91\% \\
    3-2-7-6 & 1.539\% & \textcolor{white}{0}8.33\% & \textcolor{white}{0}5.86\% \\
    3-6-2-7 & 1.894\% & 10.42\% & \textcolor{gray!20}{0}6.55\% \\
    3-6-7-2 & 1.894\% & 10.42\% & \textcolor{white}{0}6.55\% \\
    3-7-2-6 & 1.681\% & \textcolor{gray!20}{0}8.33\% & \textcolor{gray!20}{0}5.99\% \\
    3-7-6-2 & 2.841\% & 16.67\% & 12.50\% \\ \hline
    6-2-3-7 & 2.344\% & 16.67\% & \textcolor{gray!20}{0}9.16\% \\
    6-2-7-3 & 2.225\% & 16.67\% & \textcolor{white}{0}8.90\% \\
    6-3-2-7 & 1.894\% & 10.42\% & \textcolor{gray!20}{0}7.16\% \\
    6-3-7-2 & 1.894\% & 10.42\% & \textcolor{white}{0}7.16\% \\
    6-7-2-3 & 2.225\% & 16.67\% & \textcolor{gray!20}{0}8.90\% \\
    6-7-3-2 & 1.515\% & \textcolor{white}{0}8.33\% & \textcolor{white}{0}6.18\% \\ \hline
    7-2-3-6 & 0.260\% & \textcolor{gray!20}{0}1.04\% & \textcolor{gray!20}{0}1.04\% \\
    7-2-6-3 & 2.095\% & \textcolor{white}{0}9.38\% & \textcolor{white}{0}8.20\% \\
    7-3-2-6 & 1.681\% & \textcolor{gray!20}{0}8.33\% & \textcolor{gray!20}{0}5.99\% \\
    7-3-6-2 & 2.841\% & 16.67\% & 12.50\% \\
    7-6-2-3 & 2.557\% & 19.27\% & 10.24\% \\
    7-6-3-2 & 1.515\% & \textcolor{white}{0}8.33\% & \textcolor{white}{0}5.99\% \\ \bottomrule
    \end{tabularx}
\begin{tablenotes} \footnotesize
\item
\emph{Notes:} The column Maximal 8 bias shows the mean of the eight highest distortions. \\
Draw order 2-3-6-7 is the official. \\
Draw order 2-7-3-6 (or 7-2-3-6) is the recommended.
\end{tablenotes}
\end{threeparttable} 
}   
\end{table}

Table~\ref{Table9} shows the main characteristics of the distortions for the second set of groups. Again, the order of the first two pots has at most a moderated effect, for example, in the case of draw orders 2-3-6-7 and 3-2-6-7. The two best draw orders (2-7-3-6 and 7-2-3-6) contain Pots 2, 3, 6, those that have a European team, in decreasing order according to the number of European teams, as suggested by the results of Section~\ref{Sec4}. However, placing Pot 7 as third or fourth in the draw sequence has a substantial price, the average (maximal) bias is multiplied by about 7 (8). Therefore, a robust improvement can be achieved compared to the official draw order of 2-3-6-7 by moving Pot 7 ahead in the draw sequence.
In the two worst draw orders according to the average bias, Pot 2 with three European teams follows Pots 3 and 6, which is again in line with the theoretical results. The highest maximal distortion emerges for the draw order 7-6-2-3: in this case, Argentina (the only non-European team in Pot 1) and Montenegro (the only European team in Pot 3) is assigned to the same group with a probability of 50\%, but this remains only slightly more than 30\% if the Skip mechanism is used.

To summarise, the analysis of the 2019 FIBA World Cup draw yields important lessons.
First, a particular draw order of the pots might ensure uniform distribution in contrast to other draw orders.
Second, modifying the draw order can be highly efficient even if the draw remains non-uniform, by reducing the original bias up to 90\%, which greatly exceeds the potential improvement seen in Sections~\ref{Sec51} and \ref{Sec52}.
Third, the highest bias can be close to 20 percentage points even in real-world applications.

\section{Limitations} \label{Sec6}

Naturally, our research has certain limitations.
First, Section~\ref{Sec4} considers the Skip mechanism in a particular setting with one set-based prohibited triplet list and three pots. The theoretical analysis requires further simplification for both exact and numerical calculations. Nonetheless, the results of Sections~\ref{Sec41} and \ref{Sec42} may be extended to multiple pots where set $S$ contains arbitrarily many teams from pot A, but only one from all other pots.
Second, Section~\ref{Sec5} investigates three real-world case studies that are substantially more complex than the model in Section~\ref{Sec4}: there are at least four pots, as well as several set-based prohibited pair lists. Thus, the two settings are relatively far from each other, and some transitions between these extremal cases are worth examining. Artificial datasets analogous to these tournaments can be generated to obtain more robust results---however, the required number of simulations might be prohibitive since a few hundred or even thousands of iterations are clearly insufficient.
Third, the variants of the Skip mechanism are not the only viable options to guarantee transparency. Another draw procedure, the Drop mechanism, has been used in UEFA club competitions. The comparison of the Drop and Skip mechanisms offers another promising direction for future research.

\section{Conclusions} \label{Sec7}

The current paper has analysed the most popular procedure to implement a constrained draw in sports tournaments, the Skip mechanism, which transparently chooses an assignment that satisfies arbitrary restrictions. However, its transparency makes the mechanism non-uniformly distributed over the set of valid assignments. We have investigated the role of the draw order with respect to this distortion. According to our theoretical model, if no more than two teams from a given set $S$ can play in a group and there are three pots with two pots containing only one team from the set $S$, then the draw should consider the pots in decreasing order according to the number of teams from the set $S$.

The issue is hugely relevant in practice since some major sports competitions impose similar restrictions. The findings explain the optimality of the 2018 FIFA World Cup draw (Section~\ref{Sec51}). On the other hand, the bias of the draw of the European Qualifiers for the 2022 FIFA World Cup (Section~\ref{Sec52}) and the 2019 FIBA Basketball World Cup (Section~\ref{Sec53}) could have been decreased substantially by our policy recommendation.

The paper has focused on the group draw of sports tournaments. Nonetheless, the Skip mechanism may be used in any setting where equitable and fair randomisation is needed and direct monetary compensation for fairness violations remains impossible. For instance, consider a government initiative to finance startups, which is designed as follows. The startups are initially evaluated by an AI tool, ranked, and partitioned into pots according to this ranking. The startups are allocated into groups, each group containing at most one from each pot, subject to other constraints (e.g.\ each group should have a startup led by a woman CEO). Finally, the startups in a given group are assessed by a financial expert, who chooses a given number of startups qualifying for the next round and/or getting funding.

However, the stakeholders may not fully trust the government that the groups are formed fairly. This credibility issue can be mitigated by a draw using the Skip mechanism, which also helps to understand the equal treatment of startups. In this example, the startups might be replaced by research proposals, too. In addition, a number of similar situations can be devised where potential concerns about credibility are addressed by a mechanism in which all randomisations are fully transparent, possibly within a live television show similar to a national lottery or the draw of the FIFA World Cup.

Returning to group draws, since ensuring the fairness of the draw is an explicit aim of tournament organisers \citep{UEFA2019d}, they are strongly encouraged to optimise the order of the pots to reduce the unavoidable violation of randomness at almost no price.
The decision-makers likely have a substantial scope in choosing the draw order of the Skip mechanism. Although the FIFA World Cup draw always used the Skip mechanism with the ``natural'' draw order from Pot 1 to Pot $m$, the 2020/21 \citep{UEFA2020d} and 2022/23 UEFA Nations League \citep{UEFA2021i} draws followed a reversed draw order from Pot $m$ to Pot 1. The 2021 World Men's Handball Championship used the Skip mechanism with the unusual draw order of Pot 4, Pot 3, Pot 1, Pot 2 \citep{IHF2020}.
If necessary, the optimal draw order can be achieved by a simple relabelling of the pots before the draw takes place, which can be explained to the public relatively easily.

\section*{Acknowledgements}
\addcontentsline{toc}{section}{Acknowledgements}
\noindent
This paper could not have been written without \emph{my father} (also called \emph{L\'aszl\'o Csat\'o}), who has primarily coded the simulations in Python. \\
Two anonymous reviewers and eleven colleagues provided valuable remarks and suggestions on earlier drafts. \\
We are indebted to the \href{https://en.wikipedia.org/wiki/Wikipedia_community}{Wikipedia community} for summarising important details of the sports competition discussed in the paper. \\

\section*{Funding}
\addcontentsline{toc}{section}{Funding}
\noindent
The research was supported by the National Research, Development and Innovation Office under Grants Advanced 152220 and FK 145838, and the J\'anos Bolyai Research Scholarship of the Hungarian Academy of Sciences.

\bibliographystyle{apalike}
\bibliography{All_references}

\begin{thebibliography}{}

\bibitem[Atef~Yekta et~al., 2023]{AtefYektaBergmanDay2023}
Atef~Yekta, H., Bergman, D., and Day, R. (2023).
\newblock Balancing stability and efficiency in team formation as a generalized
  roommate problem.
\newblock {\em Naval Research Logistics}, 70(1):72--88.

\bibitem[Boczo{\'n} and Wilson, 2023]{BoczonWilson2023}
Boczo{\'n}, M. and Wilson, A.~J. (2023).
\newblock Goals, constraints, and transparently fair assignments: A field study
  of randomization design in the {UEFA} {C}hampions {L}eague.
\newblock {\em Management Science}, 69(6):3474--3491.

\bibitem[Cea et~al., 2020]{CeaDuranGuajardoSureSiebertZamorano2020}
Cea, S., Dur{\'a}n, G., Guajardo, M., Saur{\'e}, D., Siebert, J., and Zamorano,
  G. (2020).
\newblock An analytics approach to the {FIFA} ranking procedure and the {W}orld
  {C}up final draw.
\newblock {\em Annals of Operations Research}, 286(1-2):119--146.

\bibitem[Csat{\'o}, 2021]{Csato2021a}
Csat{\'o}, L. (2021).
\newblock {\em Tournament Design: How Operations Research Can Improve Sports
  Rules}.
\newblock Palgrave Pivots in Sports Economics. Palgrave Macmillan, Cham,
  Switzerland.

\bibitem[Csat{\'o}, 2022]{Csato2022a}
Csat{\'o}, L. (2022).
\newblock Quantifying incentive (in)compatibility: {A} case study from sports.
\newblock {\em European Journal of Operational Research}, 302(2):717--726.

\bibitem[Csat{\'o}, 2023]{Csato2023d}
Csat{\'o}, L. (2023).
\newblock Group draw with unknown qualified teams: {A} lesson from the 2022
  {FIFA} {W}orld {C}up.
\newblock {\em International Journal of Sports Science \& Coaching},
  18(2):539--551.

\bibitem[Csat{\'o}, 2025a]{Csato2025c}
Csat{\'o}, L. (2025a).
\newblock The fairness of the group draw for the {FIFA} {W}orld {C}up.
\newblock {\em International Journal of Sports Science \& Coaching},
  20(2):554--567.

\bibitem[Csat{\'o}, 2025b]{Csato2025f}
Csat{\'o}, L. (2025b).
\newblock Random matching in balanced bipartite graphs: The (un)fairness of
  draw mechanisms used in sports.
\newblock Manuscript. {DOI}:
  \href{https://doi.org/10.48550/arXiv.2303.09274}{10.48550/arXiv.2303.09274}.

\bibitem[Csat{\'o}, 2026]{Csato2026a}
Csat{\'o}, L. (2026).
\newblock How to optimise tournament draws: The case of the {FIFA} {W}orld
  {C}up.
\newblock {\em International Transactions in Operational Research}, in press.
\newblock {DOI}: \href{https://doi.org/10.1111/itor.70194}{10.1111/itor.70194}.

\bibitem[Devriesere et~al., 2025]{DevriesereCsatoGoossens2025}
Devriesere, K., Csat{\'o}, L., and Goossens, D. (2025).
\newblock Tournament design: A review from an operational research perspective.
\newblock {\em European Journal of Operational Research}, 324(1):1--21.

\bibitem[FIBA, 2019]{FIBA2019}
FIBA (2019).
\newblock Procedure for {FIBA} {B}asketball {W}orld {C}up 2019 {D}raw.
\newblock 14 March.
  \url{https://www.fiba.basketball/basketballworldcup/2019/news/procedure-for-fiba-basketball-world-cup-2019-draw}.

\bibitem[FIFA, 2017]{FIFA2017c}
FIFA (2017).
\newblock Close-up on {F}inal {D}raw procedures.
\newblock 27 November.
  \url{https://inside.fifa.com/tournaments/mens/worldcup/2018russia/news/close-up-on-final-draw-procedures-2921440}.

\bibitem[Goossens et~al., 2020]{GoossensYiVanBulck2020}
Goossens, D., Yi, X., and Van~Bulck, D. (2020).
\newblock Fairness trade-offs in sports timetabling.
\newblock In Ley, C. and Dominicy, Y., editors, {\em Science Meets Sports: When
  Statistics Are More Than Numbers}, pages 213--244. Cambridge Scholars
  Publishing, Newcastle upon Tyne, United Kingdom.

\bibitem[Guyon, 2014]{Guyon2014a}
Guyon, J. (2014).
\newblock Rethinking the {FIFA} {W}orld {C}up\textsuperscript{{TM}} final draw.
\newblock Manuscript. {DOI}:
  \href{http://dx.doi.org/10.2139/ssrn.2424376}{10.2139/ssrn.2424376}.

\bibitem[Guyon, 2015]{Guyon2015a}
Guyon, J. (2015).
\newblock Rethinking the {FIFA} {W}orld {C}up\textsuperscript{{TM}} final draw.
\newblock {\em Journal of Quantitative Analysis in Sports}, 11(3):169--182.

\bibitem[Guyon, 2018]{Guyon2018d}
Guyon, J. (2018).
\newblock Pourquoi la {C}oupe du monde est plus \'equitable cette ann\'ee.
\newblock {\em The Conversation}.
\newblock 13 June.
  \url{https://theconversation.com/pourquoi-la-coupe-du-monde-est-plus-equitable-cette-annee-97948}.

\bibitem[Guyon et~al., 2025]{GuyonBenSalemBuchholtzerTanre2025}
Guyon, J., Ben~Salem, A., Buchholtzer, T., and Tanr\'e, M. (2025).
\newblock Drawing and scheduling the {UEFA} {C}hampions {L}eague league phase.
\newblock Manuscript. {URL}:
  \url{https://papers.ssrn.com/sol3/papers.cfm?abstract_id=5413142}.

\bibitem[IHF, 2020]{IHF2020}
IHF (2020).
\newblock Egypt 2021 draw details confirmed.
\newblock 25 August.
  \url{https://www.ihf.info/media-center/news/egypt-2021-draw-details-confirmed}.

\bibitem[Jones, 1990]{Jones1990}
Jones, M.~C. (1990).
\newblock The {W}orld {C}up draw's flaws.
\newblock {\em The Mathematical Gazette}, 74(470):335--338.

\bibitem[Kendall and Lenten, 2017]{KendallLenten2017}
Kendall, G. and Lenten, L.~J.~A. (2017).
\newblock When sports rules go awry.
\newblock {\em European Journal of Operational Research}, 257(2):377--394.

\bibitem[Kiesl, 2013]{Kiesl2013}
Kiesl, H. (2013).
\newblock Match me if you can. {M}athematische {G}edanken zur
  {C}hampions-{L}eague-{A}chtelfinalauslosung.
\newblock {\em Mitteilungen der Deutschen Mathematiker-Vereinigung},
  21(2):84--88.

\bibitem[Kl{\"o}{\ss}ner and Becker, 2013]{KlossnerBecker2013}
Kl{\"o}{\ss}ner, S. and Becker, M. (2013).
\newblock Odd odds: The {UEFA} {C}hampions {L}eague {R}ound of 16 draw.
\newblock {\em Journal of Quantitative Analysis in Sports}, 9(3):249--270.

\bibitem[Kobierecki, 2022]{Kobierecki2022}
Kobierecki, M.~M. (2022).
\newblock Politics of the group draws in football. {T}he case of restricted
  team clashes.
\newblock {\em International Journal of Sport Policy and Politics},
  14(2):321--336.

\bibitem[Kondratev et~al., 2024]{KondratevIanovskiNesterov2024}
Kondratev, A.~Y., Ianovski, E., and Nesterov, A.~S. (2024).
\newblock How should we score athletes and candidates: Geometric scoring rules.
\newblock {\em Operations Research}, 72(6):2507--2525.

\bibitem[Laliena and L{\'o}pez, 2019]{LalienaLopez2019}
Laliena, P. and L{\'o}pez, F.~J. (2019).
\newblock Fair draws for group rounds in sport tournaments.
\newblock {\em International Transactions in Operational Research},
  26(2):439--457.

\bibitem[Laliena and L{\'o}pez, 2025]{LalienaLopez2025}
Laliena, P. and L{\'o}pez, F.~J. (2025).
\newblock Draw procedures for balanced 3-team group rounds in sports
  competitions.
\newblock {\em Annals of Operations Research}, 346(3):2065--2092.

\bibitem[Lapr{\' e} and Amato, 2025]{LapreAmato2025}
Lapr{\' e}, M.~A. and Amato, J.~G. (2025).
\newblock The impact of imbalanced groups in {UEFA} {E}uro 1980--2024 and
  comparison with the {FIFA} {W}orld {C}up.
\newblock {\em Journal of Quantitative Analysis in Sports}, in press.
\newblock {DOI}:
  \href{https://doi.org/10.1515/jqas-2024-0151}{10.1515/jqas-2024-0151}.

\bibitem[Lapr{\' e} and Palazzolo, 2023]{LaprePalazzolo2023}
Lapr{\' e}, M.~A. and Palazzolo, E.~M. (2023).
\newblock The evolution of seeding systems and the impact of imbalanced groups
  in {FIFA} {M}en's {W}orld {C}up tournaments 1954--2022.
\newblock {\em Journal of Quantitative Analysis in Sports}, 19(4):317--332.

\bibitem[Lenten and Kendall, 2022]{LentenKendall2022}
Lenten, L.~J.~A. and Kendall, G. (2022).
\newblock Scholarly sports: Influence of social science academe on sports rules
  and policy.
\newblock {\em Journal of the Operational Research Society}, 73(12):2591--2601.

\bibitem[Li et~al., 2025]{LiVanBulckGoossens2025}
Li, M., Van~Bulck, D., and Goossens, D. (2025).
\newblock Beyond leagues: A single incomplete round robin tournament for
  multi-league sports timetabling.
\newblock {\em European Journal of Operational Research}, 323(1):208--223.

\bibitem[Rathgeber and Rathgeber, 2007]{RathgeberRathgeber2007}
Rathgeber, A. and Rathgeber, H. (2007).
\newblock Why {G}ermany was supposed to be drawn in the group of death and why
  it escaped.
\newblock {\em Chance}, 20(2):22--24.

\bibitem[Roberts and Rosenthal, 2024]{RobertsRosenthal2024}
Roberts, G.~O. and Rosenthal, J.~S. (2024).
\newblock Football group draw probabilities and corrections.
\newblock {\em The Canadian Journal of Statistics}, 52(3):659--677.

\bibitem[UEFA, 2019]{UEFA2019d}
UEFA (2019).
\newblock How {UEFA} prepares for {C}hampions {L}eague and {E}uropa {L}eague
  draws.
\newblock 12 December.
  \url{https://www.uefa.com/uefaeuropaleague/news/025a-0e9f980ebd9f-5949171cd9e2-1000--how-uefa-prepares-for-champions-league-and-europa-league-draws/}.

\bibitem[UEFA, 2020a]{UEFA2020c}
UEFA (2020a).
\newblock {FIFA} {W}orld {C}up 2022 qualifying draw procedure.
\newblock
  \url{https://www.uefa.com/MultimediaFiles/Download/competitions/WorldCup/02/64/22/19/2642219_DOWNLOAD.pdf}.

\bibitem[UEFA, 2020b]{UEFA2020d}
UEFA (2020b).
\newblock {UEFA} {N}ations {L}eague 2020/21 -- league phase draw procedure.
\newblock
  \url{https://www.uefa.com/MultimediaFiles/Download/competitions/General/02/63/57/88/2635788_DOWNLOAD.pdf}.

\bibitem[UEFA, 2021]{UEFA2021i}
UEFA (2021).
\newblock {UEFA} {N}ations {L}eague 2022/23 -- league phase draw procedure.
\newblock
  \url{https://editorial.uefa.com/resources/026f-13c241515097-67a9c87ed1b2-1000/unl_2022-23_league_phase_draw_procedure_en.pdf}.

\bibitem[Wallace and Haigh, 2013]{WallaceHaigh2013}
Wallace, M. and Haigh, J. (2013).
\newblock Football and marriage -- and the {UEFA} draw.
\newblock {\em Significance}, 10(2):47--48.

\end{thebibliography}

\end{document}